%% file: tcom11.tex
\documentclass[journal,draftcls,onecolumn,12pt,twoside]{IEEEtranTCOM}
\normalsize

\usepackage[T1]{fontenc}
\usepackage{amsmath,amssymb,amsthm,bbm,stmaryrd}
\usepackage{tikz,balance}
\usetikzlibrary{shapes,decorations.pathreplacing}
\usepackage{pgfplots}
\usepackage{stfloats,cite}
\usepackage[colorlinks,linktocpage=true,pagebackref]{hyperref}

\newcommand{\indicator}[1]{\mathbbm{1}_{\left\{ {#1} \right\} }}
\newcommand{\entropy}[1]{H\!\left(#1\right)}
\newcommand{\sw}{SW}
\renewcommand{\sc}{spatially-coupled}
\newcommand{\x}{\mathsf{x}}

\renewcommand{\a}{\mathsf{a}}
\renewcommand{\b}{\mathsf{b}}

\providecommand{\abs}[1]{\left\lvert#1\right\rvert}
\newtheorem{theorem}{Theorem}
\newtheorem{remark}{Remark}
\newtheorem{lem}{Lemma}
\newtheorem{corollary}{Corollary}

\pgfplotsset{tick label style={
font=\Large}}

\begin{document}
\pgfdeclarelayer{background}
\pgfdeclarelayer{foreground}
\pgfsetlayers{background,main,foreground}

\title{Code Design for the Noisy Slepian-Wolf Problem}

\author{Arvind~Yedla,~\IEEEmembership{Student Member,~IEEE,}
  Henry~D.~Pfister,~\IEEEmembership{Member,~IEEE,}
  and~Krishna~R.~Narayanan,~\IEEEmembership{Senior Member,~IEEE}%
  \thanks{This work was supported in part by the National Science
    Foundation under Grant No. CCR-0515296 and by the Qatar National
    Research Foundation under its National Research Priorities
    Program. The material in this paper was presented in part at the
    47th Annual Allerton Conference on Communications, Control and
    Computing, Monticello, IL, October 2009 and in part at the 6th
    IEEE International Symposium on Turbo Codes and Related Topics
    (ISTC), Brest, France, September 2010.}%
  \thanks{The authors are with the Department of Electrical and
    Computer Engineering, Texas A\&M University, College Station, TX
    77843, USA (email: yarvind@tamu.edu; hpfister@tamu.edu;
    krn@tamu.edu).}}

\markboth{IEEE Transactions on Communications}%
{Submitted paper}
\maketitle

\begin{abstract}
  We consider a noisy Slepian-Wolf problem where two correlated
  sources are separately encoded (using codes of fixed rate) and
  transmitted over two independent binary memoryless symmetric
  channels. The capacity of each channel is characterized by a single
  parameter which is not known at the transmitter. The goal is to
  design systems that retain near-optimal performance without channel
  knowledge at the transmitter. 

  It was conjectured that it may be hard to design codes that perform
  well for symmetric channel conditions. In this work, we present a
  provable capacity-achieving sequence of LDGM ensembles for the
  erasure Slepian-Wolf problem with symmetric channel conditions. We
  also introduce a staggered structure which enables codes optimized
  for single user channels to perform well for symmetric channel
  conditions.

  We provide a generic framework for analyzing the performance of
  joint iterative decoding, using density evolution. Using
  differential evolution, we design punctured systematic LDPC codes to
  maximize the region of achievable channel conditions. The resulting
  codes are then staggered to further increase the region of
  achievable parameters. The main contribution of this paper is to
  demonstrate that properly designed irregular LDPC codes can perform
  well simultaneously over a wide range of channel parameters.
\end{abstract}


\IEEEpeerreviewmaketitle

\section{Introduction}
\label{sec:introduction}
\input{introduction}

\section{Problem Setup}
\label{sec:problem-setup}
\input{problem_setup}

\section{Analysis}
\label{sec:analysis}

\subsection{LDGM Codes}
\label{sec:ldgm-codes}
\input{ldgm_codes}

\subsection{Puncturing and LDPC Codes}
\label{sec:puncturing}
\input{puncturing}

\subsection{Density Evolution for LDPC codes}
\label{sec:density-evolution}
\input{ldpc_de}

\subsection{Staggered Block Codes}
\label{sec:single-user-codes}
\input{staggering}

\subsection{Differential Evolution}
\label{sec:diff-ev}
\input{diff_ev}

\section{Results and Concluding Remarks}
\label{sec:results}
\input{results}

\appendices
\input{appendix}

\bibliographystyle{IEEEtran}
\bibliography{IEEEabrv,WCLabrv,WCLbib,WCLnewbib}

\end{document}

%% file: introduction.tex
Wireless sensor networks have become very popular in recent years and
are being increasingly used in many commercial applications. A good
survey of the problems involved with designing sensor networks can be
found in \cite{Akyildiz-commag02,Chong-03}. A sensor network typically
has several transceivers (also called nodes), each of which has one or
several sensors. The task of these sensor nodes is to collect
measurements, encode them, and transmit them to some data collection
points. The topology of sensor networks varies widely with the
application, but typically the data from all the nodes is transmitted
to a central node, also known as a gateway node, before further
processing is done on the data. This problem is often referred to as
the sensor reachback problem. There are many constraints on the size
and cost of the networks, so the nodes have limited computational
capabilities, communication bandwidth etc.  Hence the nodes have to
perform distributed encoding, despite having to transmit correlated
data. One of the main goals in the area of wireless sensor networks is
to reduce the amount of transmitted data by taking advantage of the
correlation between the sources. In many cases, there is generally a
medium access control (MAC) protocol in place, which eliminates
interference between the different nodes. In this case, one can assume
that each node transmits through an independent channel, from the same
channel family. A simple sensor network consisting of two sensors is
shown in Fig.~\ref{fig:sys-model}. This problem of distributed
encoding and transmission over independent channels gives a noisy
version of the celebrated Slepian-Wolf (\sw{}) problem. The SW problem
was introduced and solved in the landmark paper \cite{Slepian-it73},
and shows that the optimal coding scheme suffers no loss in
performance (in terms of rate) even in the absence of communication
between the various encoders. A variety of coding schemes have been
designed that can achieve the \sw{} bound when channel state
information is known at the transmitter.

\subsection{Prior Work}
\label{sec:prior-work-1}

The first practical \sw{} coding scheme was introduced by Wyner and is
based on linear error-correcting codes \cite{Wyner-it74}. Chen et
al. related the \sw{} (distributed source coding) problem to channel
coding via an equivalent channel describing the source correlation
\cite{Chen-isit06,Delsarte-it82}. Using this observation they used
density evolution to design LDPC coset codes that approach the \sw{}
bound. Distributed source coding using syndromes (DISCUS) also
provides a practical method to transmit information for this problem
when the encoding rates are restricted to the corner points of the
rate region \cite{Pradhan-it03}.

For transmision over noisy channels, separation between source and
channel coding is known to be optimal when the channel state is known
at the transmitter \cite{Barros-it06}. When the channel state is
unknown, it is still desirable to take a joint source-channel coding
(JSCC) approach (via direct channel coding and joint decoding at the
receiver). The main reason is that separate source and channel coding
requires compression of the sources to their joint entropy prior to
channel encoding. After that, the variation in one channel's parameter
cannot be offset by variation in the other channel. Further advantages
of JSCC, over separated source coding and channel coding, are
discussed further in \cite{GarciaFrias-dcc01}.

The performance of concatenated LDGM codes has been studied in
\cite{Zhong-eurasip05} and that of Turbo codes in
\cite{GarciaFrias-dcc01}. Serially concatenated LDPC and
convolutional codes were also considered in \cite{Hu-globe04}, where
the outer LDPC code is used for distributed source coding.

It was conjectured in \cite{Martalo-ita10} that LDPC codes do not
perform well for the noisy \sw{} problem\footnote{The authors consider
  only systematic LDPC codes} and that it is hard to design codes that
perform well for symmetric channel conditions. In this work, we show a
sequence of LDGM codes which approach the \sw{} bound for symmetric
channel conditions.

\subsection{Universality}
\label{sec:universality}

Another interesting line of research in the area of sensor networks is
the sensor location problem. The sensor locations are optimized in
order to collect the most relevant data. A possibility of using moving
sensors is present in a variety of applications, including air
pollution estimation, traffic surveillance etc. \cite{Chong-03}. A
natural consequence of this is the variation in channel conditions as
a result of sensor mobility. As a result, it may be unreasonable to
assume that transmitters have detailed channel state information. This
problem of unknown channel state at the transmitter naturally arises
in the context of many multi-user scenarios, including cellular
telephony.

For fixed user code rates, reliable communication is theoretically
possible over a wide range of channel conditions \cite{Cover-1991}. We
call a system \emph{universal} if it provides good performance for all
system parameters that do not violate theoretical limits. This
designation neglects the fact that the receiver is assumed to have
channel state information and is based on the standard assumption that
the receiver can estimate the channel state with negligible pilot
overhead. While irregular LDPC codes can be optimized to approach
capacity for any particular channel condition, the performance can
deteriorate markedly as the channel conditions change.  So, we design
LDPC codes which are robust to variation in channel
conditions\footnote{Unfortunately, the LDGM codes that achieve the
  symmetric channel condition are not universal.}. Such schemes are
desirable because they minimize the outage probability for
quasi-static channels (e.g., when a probability distribution is
assigned to the set of possible channel parameters).




%% file: problem_setup.tex
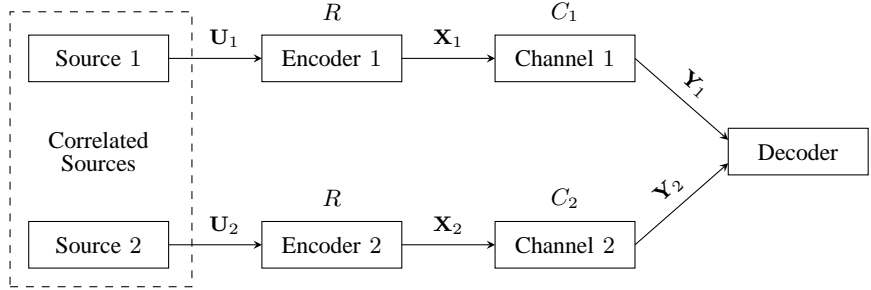
\begin{figure}
  \centering
  \input{system_model.tex}
  \caption{System Model}
  \label{fig:sys-model}
\end{figure}

Consider the problem of transmitting the outputs of two discrete
memoryless correlated sources, $\left(U_1,U_2\right)$, to a central
receiver through two independent discrete memoryless channels with
capacities $C_1$ and $C_2$, respectively. The system model is shown in
Figure~\ref{fig:sys-model}. We will assume that the channels belong to
the same channel family, and that each channel can be parametrized by a
single parameter $\alpha$ (e.g., the erasure probability
for erasure channels). The two encoders are not allowed to communicate.
Hence they must use independent encoding functions, which map $k$ input
symbols $(\mathbf{U}_1\text{ and } \mathbf{U}_2)$ to $n_1$ and $n_2$
output symbols $(\mathbf{X}_1\text{ and } \mathbf{X}_2)$,
respectively. The rates of the encoders are given by $R_1 = k/n_1$ and
$R_2 = k/n_2$. The decoder receives $(\mathbf{Y}_1,\mathbf{Y}_2)$ and
makes an estimate of $(\mathbf{U}_1,\mathbf{U}_2)$. 

The problem we consider is to design a graph-based code, for which a
joint iterative decoder can successfully decode over a large set of
channel parameters. For simplicity, we assume that both the encoders
use identical codes of rate $R$ (i.e., $R = k/n,n_1 = n_2 =
n$). Reliable transmission over a channel pair $(\alpha_1,\alpha_2)$
is possible as long as the \sw{} conditions (\ref{eq:sw}) are
satisfied.
\begin{equation}
  \label{eq:sw}
  \begin{split}
    \frac{C_1 (\alpha_1)}{R} &\geq \entropy{U_1\middle |U_2} \\
    \frac{C_2 (\alpha_2)}{R} &\geq \entropy{U_2\middle |U_1} \\
    \frac{C_1 (\alpha_1)}{R}+\frac{C_2 (\alpha_2)}{R}&\geq
    \entropy{U_1,U_2}
  \end{split}
\end{equation}

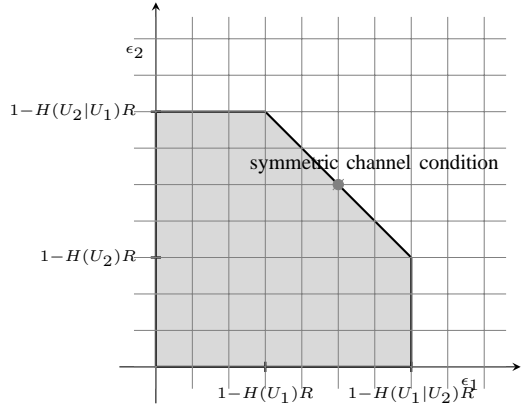
\begin{figure}
  \centering
  \input{slepian_wolf_region.tex}
  \caption{The \sw{} region for erasure channels, for a fixed
    rate pair $(R,R)$}
  \label{fig:sw}
\end{figure}

For a given pair of encoding functions of rate $R$ and a joint
decoding algorithm, a pair of channel parameters $(\alpha_1,\alpha_2)$
is \emph{achievable} if the encoder/decoder combination can achieve an
arbitrarily low probability of error for limiting block-lengths (i.e.,
$k \rightarrow \infty$). We define the achievable channel parameter
region (ACPR) as the set of all channel parameters which are
achievable. Note that the ACPR is the set of all channel parameters
for which successful recovery of the sources is possible for a fixed
encoding rate pair $(R,R)$. We also define the \emph{\sw{} region} as
the set of all channel parameters $(\alpha_1,\alpha_2)$ for which
(\ref{eq:sw}) is satisfied. The \sw{} region for the erasure channel
family is shown in Figure~\ref{fig:sw}.

In this paper, we consider the following scenarios:
\begin{enumerate}
\item The channels are erasure channels and the source correlation is
  modeled through erasures.
\item The channels are additive white Gaussian noise (AWGN) channels and
  the source correlation is modeled through a virtual correlation
  channel analogous to a binary symmetric channel (BSC).
\end{enumerate}
These models might appear restrictive, but we believe they provide
sufficient insight for the design of codes that perform well for
arbitrary correlated sources and channels. Our analysis in
Section~\ref{sec:analysis} admits general correlation models and
memoryless channels.

\subsection{Erasure Correlation}
\label{sec:erasure-correlation}

The erasure system model is based on communication over binary erasure
channels (BECs) and the source correlation is also modeled through
erasures. Let $Z$ be a Bernoulli-$p$ random variable and $X,X'$ be
i.i.d. Bernoulli-$\frac 12$ random variables. The sources $U_1$ and
$U_2$ are defined by
\begin{equation*}
  \left(U_1,U_2\right) =
  \begin{cases}
    (X,X') & \text{if } Z=0 \\
    (X,X)  & \text{if } Z=1 
  \end{cases}.
\end{equation*}
We have $\entropy{U_1|U_2} = \entropy{U_2|U_1} = 1-p$ and
$\entropy{U_1,U_2} = 2-p$. This correlation model can be incorporated
into the Tanner graph (see
Section~\ref{sec:ldgm-codes},~\ref{sec:density-evolution}) at the
decoder with the presence or absence of a check node between the
source bits depending on the auxiliary random variable $Z$. Note that
the decoder requires the realization of the random variable $Z$, for
each source bit, as side information. Because of this requirement, one
might consider this a toy model that is used mainly to gain a better
understanding of the problem. Still, a very similar model was used
recently to model internet file streaming from multiple sources
\cite{Luby-isit11}.

This model can also be thought of as having two types of BSC
correlation between the source bits (as described in the next
section), one with parameter $0$ and one with parameter $1$. The
correlation parameter $p$ determines how many bits are correlated with
parameter $1$. The receiver knows which bits are correlated with
parameter $1$.

\subsection{BSC Correlation}
\label{sec:bsc-correlation}

A more realistic model is the BSC/AWGN system model, where
communication takes place over a binary-input additive white
Gaussian-noise channel (BAWGNC) and the symmetric source correlation
is defined in terms of a single parameter, namely $p=\Pr(U_1=U_2)$. It
is useful to visualize this correlation by the presence of an
auxiliary binary symmetric channel (BSC) with parameter $1-p$ between
the sources. In other words, $U_2$ is the output of a BSC with input
$U_1$ i.e., $U_2 = U_1 + Z$. Here $Z$ is a Bernoulli-($1-p$) random
variable and can be thought of as an {\em error}. Let $h_2(\cdot)$
denote the binary entropy function. Then, $\entropy{U_1|U_2} =
\entropy{U_2|U_1} = h_2(p)$ and $\entropy{U_1,U_2} = 1 + h_2(p)$.

This correlation model can be incorporated into the Tanner graph at the
decoder (described in Section~\ref{sec:density-evolution}) as check nodes
between the source bits, with a hidden node representing the auxiliary
random variable $Z$ (which carries a constant log-likelihood ratio
$\log\frac{1-p}{p}$) attached to the check node. For this scenario, the
decoder does not require any side information i.e., it does not need to
know the realization of the auxiliary random variable $Z$.

\subsection{Existence of Universal codes}
\label{sec:exist-univ-codes}

In this section, we discuss the existence of universal coding schemes,
for the system model considered in Figure~\ref{fig:sys-model}. Let
$I_{\alpha_1}(X_1;Y_1)$ and $I_{\alpha_2}(X_2;Y_2)$ denote the mutual
information between the channel inputs and outputs when the channel
parameters are given by $\alpha_1$ and $\alpha_2$. The following
theorem shows the existence of codes which have large ACPRs.

\begin{theorem}
  \label{thm:random_binning}
  Consider encoders with rate pair $(R,R)$. For a fixed pair of
  channel conditions $(\alpha_1,\alpha_2)$, which are not known at the
  transmitter, random coding with typical-set decoding at the receiver
  can achieve an average probability of error
  $\bar{P}_{e,\alpha_1,\alpha_2}$ bounded above by
  $2^{-n\gamma(\alpha_1,\alpha_2)}$, where
    \begin{align*}
      \gamma(\alpha_1,\alpha_2) &= \min\big\{I_{\alpha_1}(X_1;Y_1)-RH(U_1\mid U_2),\\
      &\phantom{= }I_{\alpha_2}(X_2;Y_2)-RH(U_2\mid U_1),\\
      &\phantom{= }I_{\alpha_1}(X_1;Y_1) + I_{\alpha_2}(X_2;Y_2) -
      RH(U_1,U_2)\big\}.
    \end{align*}
    Hence, there exists an encoder for which the probability of error
    \begin{align*}
      P_{e,\alpha_1,\alpha_2} \leq 2^{-n\gamma(\alpha_1,\alpha_2)}.
    \end{align*}
\end{theorem}
\begin{IEEEproof}
  This follows from extending the proofs in
  \cite{Blackwell-annmathstats59} to the \sw{} problem.
\end{IEEEproof}

\begin{remark}
  A simple application of Fano's inequality shows that any pair of
  channel parameters for which $\gamma(\alpha_1,\alpha_2) < 0$
  are not achievable (the probability of error is strictly bounded away
  from zero). For binary memoryless symmetric (BMS) channels, the
  condition $\gamma(\alpha_1,\alpha_2) > 0$ translates to the
  conditions in (\ref{eq:sw}). So, the conditions in (\ref{eq:sw}) are
  both necessary and sufficient for transmission over BMS channels.
\end{remark}

\begin{remark}
  For BMS channels, the achievable channel parameter region for a
  random code is a dense subset of the entire \sw{} region for
  limiting block-lengths. This follows by using
  Theorem~\ref{thm:random_binning} and applying the Markov
  inequality. This result is also easily extended to random linear
  codes.
\end{remark}


We conclude that, for a given rate pair $(R,R)$, a single
encoder/decoder pair suffices to communicate the sources over all pairs
of BMS channels in the \sw{} region. Thus, one can obtain optimal
performance even without knowledge of $(\alpha_1,\alpha_2)$ at the
transmitter.  We refer to such encoder/decoder pairs as being {\em
  universal}. This means that random codes with typical-set decoding are
universal for BMS channels.


While random codes with typical-set decoding are universally good,
encoding and decoding is known to be impractical due to its large
complexity. This motivates the search for low complexity
encoding/decoding schemes which are universal.


%% file: system_model.tex
\begin{tikzpicture}[scale=0.62,>=stealth,xshift=-1cm]
\draw (0,0) rectangle +(3,1);
\draw (1.5,.5) node {\footnotesize Source $2$};
\draw (0,4) rectangle +(3,1);
\draw (1.5,4.5) node {\footnotesize Source $1$};
\draw (5,0) rectangle +(3,1);
\draw (6.5,.5) node {\footnotesize Encoder $2$};
\draw (6.5,1.5) node {\footnotesize $R$};
\draw (5,4) rectangle +(3,1);
\draw (6.5,4.5) node {\footnotesize Encoder $1$};
\draw (6.5,5.5) node {\footnotesize $R$};
\draw (10,0) rectangle +(3,1);
\draw (11.5,.5) node {\footnotesize Channel $2$};
\draw (11.5,1.5) node {\footnotesize $C_2$};
\draw (10,4) rectangle +(3,1);
\draw (11.5,4.5) node {\footnotesize Channel $1$};
\draw (11.5,5.5) node {\footnotesize $C_1$};
\draw (15,2) rectangle +(3,1);
\draw (16.5,2.5) node {\footnotesize Decoder};

\draw[->] (3,0.5) -- (5,0.5) node[pos=0.6,above] {$\scriptstyle{\mathbf{U}}_2$};
\draw[->] (3,4.5) -- (5,4.5) node[pos=0.6,above] {$\scriptstyle{\mathbf{U}}_1$};
\draw[->] (8,0.5) -- (10,0.5) node[midway,above] {$\scriptstyle{\mathbf{X}}_2$};
\draw[->] (8,4.5) -- (10,4.5) node[midway,above] {$\scriptstyle{\mathbf{X}}_1$};
\draw[->] (13,0.5) -- (15,2.25) node[sloped,midway,above] {$\scriptstyle{\mathbf{Y}}_2$};
\draw[->] (13,4.5) -- (15,2.75) node[sloped,midway,above] {$\scriptstyle{\mathbf{Y}}_1$};
\draw[dashed] (-0.4,-0.4) rectangle +(3.9,5.9);
\draw (1.5,2.75) node {\footnotesize Correlated};
\draw (1.5,2.25) node {\footnotesize Sources};
\end{tikzpicture}


%% file: slepian_wolf_region.tex
\begin{tikzpicture}[>=stealth,scale=0.97]
\draw[->] (-0.5,0) -- (5,0) node[very near end,sloped,below] {$\scriptscriptstyle\epsilon_1$};
\draw[->] (0,-0.5) -- (0,5) node[very near end,left] {$\scriptscriptstyle\epsilon_2$};
\shade[top color=gray!30!white, bottom color=gray!30!white] (0,0) -- (0,3.5) -- (1.5,3.5) -- (3.5,1.5) -- (3.5,0) -- cycle;
\draw[thick] (0,0) -- (0,3.5) -- (1.5,3.5) -- (3.5,1.5) -- (3.5,0) -- cycle;
\draw[gray, very thin] (2.425,2.425) -- (2.575,2.575);
\draw[thick] (3.5cm,-2pt) -- (3.5cm,2pt) node[below=5pt] {$\scriptscriptstyle 1-H(U_1\mid U_2)R$};
\draw[thick] (1.5cm,-2pt) -- (1.5cm,2pt) node[below=5pt] {$\scriptscriptstyle{1-H(U_1)R}$};
\draw[thick] (-2pt,3.5cm) -- (2pt,3.5cm) node[left=5pt] {$\scriptscriptstyle 1-H(U_2\mid U_1)R$};
\draw[thick] (-2pt,1.5cm) -- (2pt,1.5cm) node[left=5pt] {$\scriptscriptstyle{1-H(U_2)R}$};
\draw[step=.5cm,gray,very thin] (-0.3,-0.3) grid (4.8,4.8);
\draw (3,2.5) node[above] {\scriptsize symmetric channel condition};
\filldraw[gray] (2.5,2.5)
circle (2pt);
\end{tikzpicture}


%% file: ldgm_codes.tex
Assume that the sequences $\mathbf{U}_1$ and $\mathbf{U}_2$ are
encoded using LDGM codes with a degree distribution pair
$\left(\lambda,\rho\right)$. Based on standard notation
\cite{RU-2008}, we let $\lambda(x) = \sum_i \lambda_i x^{i-1}$ be the
degree distribution (from an edge perspective) corresponding to the
variable nodes and $\rho(x) = \sum_i \rho_i x^{i-1}$ be the degree
distribution (from an edge perspective) of the parity-check nodes in
the decoding graph. The coefficient $\lambda_i$ (resp. $\rho_i$) gives
the fraction of edges that connect to the variable nodes
(resp. parity-check nodes) of degree $i$. Likewise, $L_i$
(resp. $R_i$) is the fraction of variable nodes (resp. check nodes)
with degree $i$.

Since the encoded variable nodes are are attached to the check nodes
randomly, the degree of each variable node is a Poisson random
variable whose mean is given by the average number of edges attached
to each check node. This mean is given by $\mathsf{m} = R'(1)$, where
$R'(1)$ is the average check degree. Therefore, the resulting degree
distribution is $L(x) = \text{e}^{\mathsf{m}(x-1)}$. Throughout this
section, we consider the erasure correlation model described in
Section~\ref{sec:erasure-correlation}.

\begin{figure}
  \centering
  \input{tanner_ldgm}
  \caption{Tanner Graph of an LDGM (LT) Code with erasure
    correlation between the sources}
  \label{fig:tanner}
\end{figure}
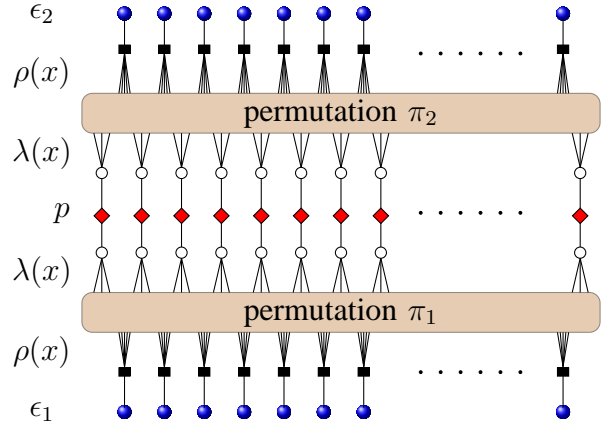

The Tanner graph \cite{RU-2008} for the code is shown in
Fig. \ref{fig:tanner}. Code $1$ corresponds to the bottom half of the
graph, code $2$ corresponds to the top half and both the codes are
connected by correlation nodes at the source variable nodes. One can
verify that the computation graph for decoding a particular bit is
asymptotically tree-like, for a fixed number of iterations as the
blocklength tends to infinity. This enables the use of density
evolution to compute the performance of the joint iterative decoder.

Let $x_{\ell}$ and $y_{\ell}$ denote the average erasure probability
of the variable nodes at iteration $\ell$ for users $1$ and $2$
respectively. The density evolution equations \cite{RU-2008} in terms
of the variable-node to check-node messages can be written as
\begin{align*}
    x_{\ell+1} &= \left[(1-p)+pL\left(\varrho(\epsilon_2,y_{\ell})\right)\right]\lambda\left(\varrho(\epsilon_1,x_{\ell})\right)\\
    y_{\ell+1} &= \left[(1-p)+pL\left(\varrho(\epsilon_1,x_{\ell})\right)\right]\lambda\left(\varrho(\epsilon_2,y_{\ell})\right),
\end{align*}
where $\varrho(\epsilon,x) = 1 - (1-\epsilon)\rho(1-x)$.  Notice that,
for LT codes, the variable-node degree distribution from the edge
perspective is given by $\lambda^{(i)}(x) = L^{(i)}(x)$ because
$\lambda(x) \triangleq L'(x)/L'(1) = L(x)$, when $L(x)$ is
Poisson. With this simplification, the density evolution for symmetric
channel conditions ($\epsilon_1=\epsilon_2=\epsilon$) can be written
as
\begin{align}
  \label{eq:one-de} 
  x_{\ell+1} \! = \! \bigl[(1-p)+p\lambda\bigl(1-(1-\epsilon)\rho(1-x_{\ell})\bigr)\bigr]\lambda\bigl(1-(1-\epsilon)\rho(1-x_{\ell})\bigr).
\end{align}
This recursion can be solved analytically, resulting in the unique
non-negative $\rho(x)$ which satisfies
\begin{align*}
  x = \bigl[(1-p)+p\lambda\bigl(1-(1-\epsilon)\rho(1-x)\bigr)\bigr]\lambda\bigl(1-(1-\epsilon)\rho(1-x)\bigr).
\end{align*}
The solution is given by
\begin{align*}
  \rho(x) &= \frac{-1}{\alpha(1-\epsilon)}\cdot\log\left(\frac{\sqrt{(1-p)^2+4p(1-x)}-(1-p)}{2p}\right)\\
  &= \frac{1}{\alpha(1-\epsilon)}\sum_{i=1}^{\infty}
  \frac{\sum_{k=0}^{i-1}\binom{2i-1}{k}p^k}{i(1+p)^{2i-1}}x^i,
\end{align*}
which is not a valid degree distribution because it has infinite
mean. To overcome this, we define a truncated version of the check
degree distribution via
\begin{equation}
  \label{eq:gnp_def}
  \begin{split}
    \rho^N(x) &= \frac{\mu+\sum_{i=1}^{N} \frac{\sum_{k=0}^{i-1}\binom{2i-1}{k}p^k}{i(1+p)^{2i-1}}x^i+x^N}{\mu+G_N(p)+1}\\
    G_N(p) &=
    \sum_{i=1}^N\frac{\sum_{k=0}^{i-1}\binom{2i-1}{k}p^k}{i(1+p)^{2i-1}},
  \end{split}
\end{equation}
for some $\mu>0$ and $N\in \mathbb{N}$.  This is a well defined degree
distribution as all the coefficients are non-negative and
$\rho^N(1)=1$. The parameter $\mu$ increases the number of degree one
generator nodes and is introduced in order to overcome the stability
problem at the beginning of the decoding process
\cite{Luby-focs02}.

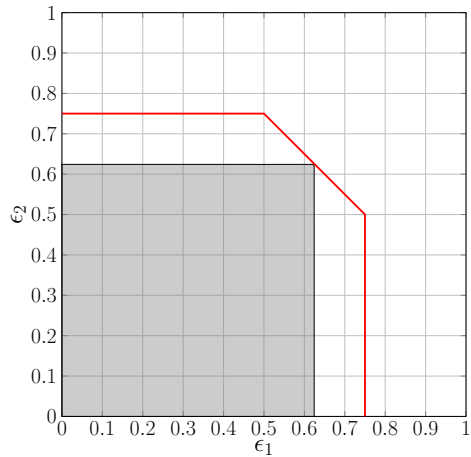
\begin{figure}[b!]
  \centering
  \input{erasure_opt_lt_roc}
  \vspace{-3mm}
  \caption{ACPR (Density Evolution threshold) of the optimized (erasure
    channel) LT Code with $N=2048$}
  \label{fig:roc_ldpc_erasure_lt}
\end{figure}

\begin{theorem}
  \label{thm-sum-rate}
  Consider transmission over erasure channels with parameters
  $\epsilon_1 = \epsilon_2 = \epsilon$. For $N\in \mathbb{N}$ and
  $\mu>0$, define
  \begin{align*}
    G_N(p) =
    \sum_{i=1}^N\frac{\sum_{k=0}^{i-1}\binom{2i-1}{k}p^k}{i(1+p)^{2i-1}},\text{
    and } \mathsf{m} = \frac{\mu + G_N(p) + 1}{1 -\epsilon}.
  \end{align*}
  Then, in the limit of infinite blocklengths, the ensemble
  LDGM$\left(n,\lambda(x),\rho^N(x)\right)$, where
  \begin{equation}
    \label{eq:ensemble}
      \lambda(x) = \text{e}^{\mathsf{m}(x-1)}\text{ and }
      \rho^N(x) = \frac{\mu+\sum_{i=1}^{N} \frac{\sum_{k=0}^{i-1}\binom{2i-1}{k}p^k}{i(1+p)^{2i-1}}x^i+x^N}{\mu+G_N(p)+1},
  \end{equation}
  enables transmission at a rate $R =
  \frac{(1-\epsilon)(1-\text{e}^{-\mathsf{m}})}{\mu+1-p/2}$, with a bit
  error probability not exceeding $1/N$.
\end{theorem}
\begin{IEEEproof}
  See Appendix~\ref{sec:proof-theorem-refthm}.
\end{IEEEproof}
From Theorem \ref{thm-sum-rate}, we conclude that the optimized
ensemble LDGM$\left(n,\lambda(x),\rho^N(x)\right)$ can achieve the
extremal symmetric point of the capacity region. Unfortunately, one
can show that this ensemble cannot simultaneously achieve both the
extremal symmetric point and the corner points of the SW region. In
Figure~\ref{fig:roc_ldpc_erasure_lt}, this can also be observed
numerically via the density evolution ACPR (DE-ACPR) of this ensemble
for $N=2048$.

%% file: tanner_ldgm.tex
\begin{tikzpicture}[scale=0.53]
\draw[rounded corners,fill=brown,opacity=0.4] (-1,2) rectangle +(13,1);
\draw (5.5,2.5) node {permutation $\pi_1$};
\draw[rounded corners,fill=brown,opacity=0.4] (-1,7) rectangle +(13,1);
\draw (5.5,7.5) node {permutation $\pi_2$};
\foreach \x in {0,1,2,3,4,5,6,11}
{
\shade[ball color=blue] (\x,0)+(2pt,0pt) circle (5pt);
\draw (\x,0)+(2pt,3pt) -- ([xshift=2pt]\x,1); 
\filldraw[black] (\x,0.9)+(-2pt,0pt) rectangle +(6pt,6pt);
\draw (\x,1)+(2pt,4pt) -- (\x,2);
\draw (\x,1)+(2pt,4pt) -- ([xshift = 4pt]\x,2);
\draw (\x,1)+(2pt,4pt) -- ([xshift = 2pt]\x,2);
\draw (\x,1)+(2pt,4pt) -- ([xshift = -2pt]\x,2);
\draw (\x,1)+(2pt,4pt) -- ([xshift = 6pt]\x,2);
\draw (\x,4)+(0.5,0) circle (4pt);
\draw (\x,4)+(0.5cm,-4pt) -- ([xshift=0.5cm]\x,3);
\draw (\x,4)+(0.5cm,-4pt) -- ([xshift=0.3cm]\x,3);
\draw (\x,4)+(0.5cm,-4pt) -- ([xshift=0.7cm]\x,3);

\draw (\x,4)+(0.5cm,4pt) -- ([xshift=0.5cm]\x,5.85);

\draw (\x,6)+(0.5,0) circle (4pt);
\draw (\x,6)+(0.5cm,4pt) -- ([xshift=0.5cm]\x,7);
\draw (\x,6)+(0.5cm,4pt) -- ([xshift=0.3cm]\x,7);
\draw (\x,6)+(0.5cm,4pt) -- ([xshift=0.7cm]\x,7);
\draw (\x,9)+(2pt,0pt) -- (\x,8);
\draw (\x,9)+(2pt,0pt) -- ([xshift = 4pt]\x,8);
\draw (\x,9)+(2pt,0pt) -- ([xshift = 2pt]\x,8);
\draw (\x,9)+(2pt,0pt) -- ([xshift = -2pt]\x,8);
\draw (\x,9)+(2pt,0pt) -- ([xshift = 6pt]\x,8);
\shade[ball color=blue] (\x,10)+(2pt,0pt) circle (5pt);
\draw (\x,10)+(2pt,-3pt) -- ([xshift=2pt]\x,9.15); 
\filldraw[black] (\x,9)+(-2pt,0pt) rectangle +(6pt,6pt);
}
\draw (-0.5,4) circle (4pt);
\draw (-0.5,4)+(0cm,-4pt) -- ([xshift=0.5cm]-1,3);
\draw (-0.5,4)+(0cm,-4pt) -- ([xshift=0.3cm]-1,3);
\draw (-0.5,4)+(0cm,-4pt) -- ([xshift=0.7cm]-1,3);

\draw (-1,4)+(0.5cm,4pt) -- ([xshift=0.5cm]-1,5.85);

\draw (-0.5,6) circle (4pt);
\draw (-0.5,6)+(0cm,4pt) -- ([xshift=0.5cm]-1,7);
\draw (-0.5,6)+(0cm,4pt) -- ([xshift=0.3cm]-1,7);
\draw (-0.5,6)+(0cm,4pt) -- ([xshift=0.7cm]-1,7);
\foreach \x in {0.5,1.5,2.5,...,6.5,11.5} {
\node[diamond,draw=black,fill=red,inner sep=0pt,minimum size=6pt] at (\x,4.925) {};
}
\node[diamond,draw=black,fill=red,inner sep=0pt,minimum size=6pt] at (-0.5,4.925) {};

\foreach \x in {7.5,8,8.5,...,10} {
\foreach \y in {1,5,9} {
\filldraw (\x,\y) circle (1pt);
}}
\draw (-2,0) node {$\epsilon_1$};
\draw (-2,1.5) node {$\rho(x)$};
\draw (-2,3.5) node {$\lambda(x)$};
\draw (-1.5,5) node {$p$};
\draw (-2,10) node {$\epsilon_2$};
\draw (-2,8.5) node {$\rho(x)$};
\draw (-2,6.5) node {$\lambda(x)$};
\end{tikzpicture}


%% file: erasure_opt_lt_roc.tex
%

\begin{tikzpicture}[scale=0.47]

\begin{axis}[
scale only axis,
width=4.5in,
height=4.5in,
xmin=0, xmax=1,
ymin=0, ymax=1,
xtick={0,0.1,0.2,0.3,0.4,0.5,0.6,0.7,0.8,0.9,1},
ytick={0,0.1,0.2,0.3,0.4,0.5,0.6,0.7,0.8,0.9,1},
xlabel={\LARGE$\epsilon_1$},
ylabel={\LARGE$\epsilon_2$},
xmajorgrids,
ymajorgrids]

\addplot
[black,fill=gray,fill opacity=0.4]
coordinates{
 (0,0.624268)
 (0.208089,0.624268)
 (0.208437,0.624268)
 (0.208785,0.624268)
 (0.209135,0.624268)
 (0.209486,0.624268)
 (0.209838,0.624268)
 (0.210191,0.624268)
 (0.210546,0.624268)
 (0.210901,0.624268)
 (0.211258,0.624268)
 (0.211616,0.624268)
 (0.211975,0.624268)
 (0.212336,0.624268)
 (0.212698,0.624268)
 (0.213061,0.624268)
 (0.213425,0.624268)
 (0.21379,0.624268)
 (0.214157,0.624268)
 (0.214525,0.624268)
 (0.214894,0.624268)
 (0.215265,0.624268)
 (0.215636,0.624268)
 (0.21601,0.624268)
 (0.216384,0.624268)
 (0.21676,0.624268)
 (0.217137,0.624268)
 (0.217515,0.624268)
 (0.217894,0.624268)
 (0.218275,0.624268)
 (0.218658,0.624268)
 (0.219041,0.624268)
 (0.219426,0.624268)
 (0.219813,0.624268)
 (0.2202,0.624268)
 (0.220589,0.624268)
 (0.22098,0.624268)
 (0.221371,0.624268)
 (0.221765,0.624268)
 (0.222159,0.624268)
 (0.222555,0.624268)
 (0.222953,0.624268)
 (0.223352,0.624268)
 (0.223752,0.624268)
 (0.224154,0.624268)
 (0.224557,0.624268)
 (0.224961,0.624268)
 (0.225367,0.624268)
 (0.225775,0.624268)
 (0.226184,0.624268)
 (0.226594,0.624268)
 (0.227006,0.624268)
 (0.22742,0.624268)
 (0.227835,0.624268)
 (0.228251,0.624268)
 (0.228669,0.624268)
 (0.229089,0.624268)
 (0.22951,0.624268)
 (0.229933,0.624268)
 (0.230357,0.624268)
 (0.230783,0.624268)
 (0.23121,0.624268)
 (0.231639,0.624268)
 (0.23207,0.624268)
 (0.232502,0.624268)
 (0.232936,0.624268)
 (0.233371,0.624268)
 (0.233808,0.624268)
 (0.234247,0.624268)
 (0.234687,0.624268)
 (0.235129,0.624268)
 (0.235573,0.624268)
 (0.236018,0.624268)
 (0.236465,0.624268)
 (0.236914,0.624268)
 (0.237364,0.624268)
 (0.237816,0.624268)
 (0.23827,0.624268)
 (0.238726,0.624268)
 (0.239183,0.624268)
 (0.239642,0.624268)
 (0.240103,0.624268)
 (0.240566,0.624268)
 (0.24103,0.624268)
 (0.241496,0.624268)
 (0.241964,0.624268)
 (0.242434,0.624268)
 (0.242906,0.624268)
 (0.243379,0.624268)
 (0.243855,0.624268)
 (0.244332,0.624268)
 (0.244811,0.624268)
 (0.245292,0.624268)
 (0.245775,0.624268)
 (0.246259,0.624268)
 (0.246746,0.624268)
 (0.247235,0.624268)
 (0.247725,0.624268)
 (0.248218,0.624268)
 (0.248712,0.624268)
 (0.249209,0.624268)
 (0.249707,0.624268)
 (0.250207,0.624268)
 (0.25071,0.624268)
 (0.251214,0.624268)
 (0.251721,0.624268)
 (0.252229,0.624268)
 (0.25274,0.624268)
 (0.253253,0.624268)
 (0.253767,0.624268)
 (0.254284,0.624268)
 (0.254803,0.624268)
 (0.255324,0.624268)
 (0.255848,0.624268)
 (0.256373,0.624268)
 (0.2569,0.624268)
 (0.25743,0.624268)
 (0.257962,0.624268)
 (0.258496,0.624268)
 (0.259032,0.624268)
 (0.259571,0.624268)
 (0.260112,0.624268)
 (0.260655,0.624268)
 (0.2612,0.624268)
 (0.261748,0.624268)
 (0.262298,0.624268)
 (0.26285,0.624268)
 (0.263404,0.624268)
 (0.263961,0.624268)
 (0.26452,0.624268)
 (0.265082,0.624268)
 (0.265646,0.624268)
 (0.266212,0.624268)
 (0.266781,0.624268)
 (0.267352,0.624268)
 (0.267926,0.624268)
 (0.268502,0.624268)
 (0.269081,0.624268)
 (0.269662,0.624268)
 (0.270246,0.624268)
 (0.270832,0.624268)
 (0.271421,0.624268)
 (0.272012,0.624268)
 (0.272606,0.624268)
 (0.273203,0.624268)
 (0.273802,0.624268)
 (0.274403,0.624268)
 (0.275008,0.624268)
 (0.275615,0.624268)
 (0.276225,0.624268)
 (0.276837,0.624268)
 (0.277452,0.624268)
 (0.27807,0.624268)
 (0.278691,0.624268)
 (0.279314,0.624268)
 (0.279941,0.624268)
 (0.28057,0.624268)
 (0.281202,0.624268)
 (0.281836,0.624268)
 (0.282474,0.624268)
 (0.283115,0.624268)
 (0.283758,0.624268)
 (0.284404,0.624268)
 (0.285054,0.624268)
 (0.285706,0.624268)
 (0.286361,0.624268)
 (0.28702,0.624268)
 (0.287681,0.624268)
 (0.288345,0.624268)
 (0.289013,0.624268)
 (0.289683,0.624268)
 (0.290357,0.624268)
 (0.291034,0.624268)
 (0.291714,0.624268)
 (0.292397,0.624268)
 (0.293083,0.624268)
 (0.293773,0.624268)
 (0.294466,0.624268)
 (0.295162,0.624268)
 (0.295862,0.624268)
 (0.296564,0.624268)
 (0.29727,0.624268)
 (0.29798,0.624268)
 (0.298693,0.624268)
 (0.299409,0.624268)
 (0.300129,0.624268)
 (0.300852,0.624268)
 (0.301579,0.624268)
 (0.302309,0.624268)
 (0.303043,0.624268)
 (0.30378,0.624268)
 (0.304521,0.624268)
 (0.305265,0.624268)
 (0.306014,0.624268)
 (0.306766,0.624268)
 (0.307521,0.624268)
 (0.30828,0.624268)
 (0.309043,0.624268)
 (0.30981,0.624268)
 (0.310581,0.624268)
 (0.311356,0.624268)
 (0.312134,0.624268)
 (0.312916,0.624268)
 (0.313702,0.624268)
 (0.314493,0.624268)
 (0.315287,0.624268)
 (0.316085,0.624268)
 (0.316887,0.624268)
 (0.317694,0.624268)
 (0.318504,0.624268)
 (0.319319,0.624268)
 (0.320137,0.624268)
 (0.32096,0.624268)
 (0.321788,0.624268)
 (0.322619,0.624268)
 (0.323455,0.624268)
 (0.324295,0.624268)
 (0.325139,0.624268)
 (0.325988,0.624268)
 (0.326842,0.624268)
 (0.3277,0.624268)
 (0.328562,0.624268)
 (0.329429,0.624268)
 (0.3303,0.624268)
 (0.331177,0.624268)
 (0.332057,0.624268)
 (0.332943,0.624268)
 (0.333833,0.624268)
 (0.334728,0.624268)
 (0.335628,0.624268)
 (0.336533,0.624268)
 (0.337442,0.624268)
 (0.338357,0.624268)
 (0.339276,0.624268)
 (0.3402,0.624268)
 (0.34113,0.624268)
 (0.342065,0.624268)
 (0.343004,0.624268)
 (0.343949,0.624268)
 (0.344899,0.624268)
 (0.345855,0.624268)
 (0.346815,0.624268)
 (0.347782,0.624268)
 (0.348753,0.624268)
 (0.34973,0.624268)
 (0.350712,0.624268)
 (0.3517,0.624268)
 (0.352694,0.624268)
 (0.353693,0.624268)
 (0.354698,0.624268)
 (0.355708,0.624268)
 (0.356724,0.624268)
 (0.357747,0.624268)
 (0.358775,0.624268)
 (0.359809,0.624268)
 (0.360848,0.624268)
 (0.361894,0.624268)
 (0.362946,0.624268)
 (0.364005,0.624268)
 (0.365069,0.624268)
 (0.366139,0.624268)
 (0.367216,0.624268)
 (0.3683,0.624268)
 (0.369389,0.624268)
 (0.370485,0.624268)
 (0.371588,0.624268)
 (0.372697,0.624268)
 (0.373813,0.624268)
 (0.374936,0.624268)
 (0.376065,0.624268)
 (0.377201,0.624268)
 (0.378344,0.624268)
 (0.379494,0.624268)
 (0.380651,0.624268)
 (0.381815,0.624268)
 (0.382986,0.624268)
 (0.384165,0.624268)
 (0.385351,0.624268)
 (0.386544,0.624268)
 (0.387744,0.624268)
 (0.388952,0.624268)
 (0.390167,0.624268)
 (0.39139,0.624268)
 (0.392621,0.624268)
 (0.39386,0.624268)
 (0.395106,0.624268)
 (0.396361,0.624268)
 (0.397623,0.624268)
 (0.398893,0.624268)
 (0.400172,0.624268)
 (0.401458,0.624268)
 (0.402753,0.624268)
 (0.404057,0.624268)
 (0.405369,0.624268)
 (0.406689,0.624268)
 (0.408018,0.624268)
 (0.409356,0.624268)
 (0.410703,0.624268)
 (0.412058,0.624268)
 (0.413422,0.624268)
 (0.414796,0.624268)
 (0.416179,0.624268)
 (0.41757,0.624268)
 (0.418972,0.624268)
 (0.420382,0.624268)
 (0.421802,0.624268)
 (0.423232,0.624268)
 (0.424672,0.624268)
 (0.426121,0.624268)
 (0.427581,0.624268)
 (0.42905,0.624268)
 (0.430529,0.624268)
 (0.432019,0.624268)
 (0.433519,0.624268)
 (0.43503,0.624268)
 (0.436551,0.624268)
 (0.438083,0.624268)
 (0.439625,0.624268)
 (0.441179,0.624268)
 (0.442743,0.624268)
 (0.444319,0.624268)
 (0.445905,0.624268)
 (0.447504,0.624268)
 (0.449113,0.624268)
 (0.450735,0.624268)
 (0.452368,0.624268)
 (0.454013,0.624268)
 (0.45567,0.624268)
 (0.457339,0.624268)
 (0.45902,0.624268)
 (0.460714,0.624268)
 (0.46242,0.624268)
 (0.46414,0.624268)
 (0.465871,0.624268)
 (0.467616,0.624268)
 (0.469374,0.624268)
 (0.471145,0.624268)
 (0.47293,0.624268)
 (0.474728,0.624268)
 (0.47654,0.624268)
 (0.478366,0.624268)
 (0.480206,0.624268)
 (0.48206,0.624268)
 (0.483834,0.624146)
 (0.485716,0.624146)
 (0.487614,0.624146)
 (0.489526,0.624146)
 (0.491453,0.624146)
 (0.493396,0.624146)
 (0.495354,0.624146)
 (0.497327,0.624146)
 (0.499316,0.624146)
 (0.501322,0.624146)
 (0.503343,0.624146)
 (0.505381,0.624146)
 (0.507435,0.624146)
 (0.509507,0.624146)
 (0.511645,0.624207)
 (0.51375,0.624207)
 (0.515873,0.624207)
 (0.518014,0.624207)
 (0.520172,0.624207)
 (0.522349,0.624207)
 (0.524543,0.624207)
 (0.526757,0.624207)
 (0.528989,0.624207)
 (0.53124,0.624207)
 (0.53351,0.624207)
 (0.5358,0.624207)
 (0.538109,0.624207)
 (0.540439,0.624207)
 (0.542788,0.624207)
 (0.545159,0.624207)
 (0.54755,0.624207)
 (0.549962,0.624207)
 (0.552395,0.624207)
 (0.55485,0.624207)
 (0.557327,0.624207)
 (0.559826,0.624207)
 (0.562348,0.624207)
 (0.564893,0.624207)
 (0.56746,0.624207)
 (0.570052,0.624207)
 (0.572667,0.624207)
 (0.575306,0.624207)
 (0.577969,0.624207)
 (0.580657,0.624207)
 (0.583371,0.624207)
 (0.586109,0.624207)
 (0.588874,0.624207)
 (0.591665,0.624207)
 (0.594482,0.624207)
 (0.597327,0.624207)
 (0.600199,0.624207)
 (0.603098,0.624207)
 (0.606026,0.624207)
 (0.608982,0.624207)
 (0.611967,0.624207)
 (0.614982,0.624207)
 (0.618026,0.624207)
 (0.620858,0.623962)
 (0.622742,0.622742)
 (0.622742,0.622742)
 (0.623962,0.620858)
 (0.624207,0.618026)
 (0.624207,0.614982)
 (0.624207,0.611967)
 (0.624207,0.608982)
 (0.624207,0.606026)
 (0.624207,0.603098)
 (0.624207,0.600199)
 (0.624207,0.597327)
 (0.624207,0.594482)
 (0.624207,0.591665)
 (0.624207,0.588874)
 (0.624207,0.586109)
 (0.624207,0.583371)
 (0.624207,0.580657)
 (0.624207,0.577969)
 (0.624207,0.575306)
 (0.624207,0.572667)
 (0.624207,0.570052)
 (0.624207,0.56746)
 (0.624207,0.564893)
 (0.624207,0.562348)
 (0.624207,0.559826)
 (0.624207,0.557327)
 (0.624207,0.55485)
 (0.624207,0.552395)
 (0.624207,0.549962)
 (0.624207,0.54755)
 (0.624207,0.545159)
 (0.624207,0.542788)
 (0.624207,0.540439)
 (0.624207,0.538109)
 (0.624207,0.5358)
 (0.624207,0.53351)
 (0.624207,0.53124)
 (0.624207,0.528989)
 (0.624207,0.526757)
 (0.624207,0.524543)
 (0.624207,0.522349)
 (0.624207,0.520172)
 (0.624207,0.518014)
 (0.624207,0.515873)
 (0.624207,0.51375)
 (0.624207,0.511645)
 (0.624146,0.509507)
 (0.624146,0.507435)
 (0.624146,0.505381)
 (0.624146,0.503343)
 (0.624146,0.501322)
 (0.624146,0.499316)
 (0.624146,0.497327)
 (0.624146,0.495354)
 (0.624146,0.493396)
 (0.624146,0.491453)
 (0.624146,0.489526)
 (0.624146,0.487614)
 (0.624146,0.485716)
 (0.624146,0.483834)
 (0.624268,0.48206)
 (0.624268,0.480206)
 (0.624268,0.478366)
 (0.624268,0.47654)
 (0.624268,0.474728)
 (0.624268,0.47293)
 (0.624268,0.471145)
 (0.624268,0.469374)
 (0.624268,0.467616)
 (0.624268,0.465871)
 (0.624268,0.46414)
 (0.624268,0.46242)
 (0.624268,0.460714)
 (0.624268,0.45902)
 (0.624268,0.457339)
 (0.624268,0.45567)
 (0.624268,0.454013)
 (0.624268,0.452368)
 (0.624268,0.450735)
 (0.624268,0.449113)
 (0.624268,0.447504)
 (0.624268,0.445905)
 (0.624268,0.444319)
 (0.624268,0.442743)
 (0.624268,0.441179)
 (0.624268,0.439625)
 (0.624268,0.438083)
 (0.624268,0.436551)
 (0.624268,0.43503)
 (0.624268,0.433519)
 (0.624268,0.432019)
 (0.624268,0.430529)
 (0.624268,0.42905)
 (0.624268,0.427581)
 (0.624268,0.426121)
 (0.624268,0.424672)
 (0.624268,0.423232)
 (0.624268,0.421802)
 (0.624268,0.420382)
 (0.624268,0.418972)
 (0.624268,0.41757)
 (0.624268,0.416179)
 (0.624268,0.414796)
 (0.624268,0.413422)
 (0.624268,0.412058)
 (0.624268,0.410703)
 (0.624268,0.409356)
 (0.624268,0.408018)
 (0.624268,0.406689)
 (0.624268,0.405369)
 (0.624268,0.404057)
 (0.624268,0.402753)
 (0.624268,0.401458)
 (0.624268,0.400172)
 (0.624268,0.398893)
 (0.624268,0.397623)
 (0.624268,0.396361)
 (0.624268,0.395106)
 (0.624268,0.39386)
 (0.624268,0.392621)
 (0.624268,0.39139)
 (0.624268,0.390167)
 (0.624268,0.388952)
 (0.624268,0.387744)
 (0.624268,0.386544)
 (0.624268,0.385351)
 (0.624268,0.384165)
 (0.624268,0.382986)
 (0.624268,0.381815)
 (0.624268,0.380651)
 (0.624268,0.379494)
 (0.624268,0.378344)
 (0.624268,0.377201)
 (0.624268,0.376065)
 (0.624268,0.374936)
 (0.624268,0.373813)
 (0.624268,0.372697)
 (0.624268,0.371588)
 (0.624268,0.370485)
 (0.624268,0.369389)
 (0.624268,0.3683)
 (0.624268,0.367216)
 (0.624268,0.366139)
 (0.624268,0.365069)
 (0.624268,0.364005)
 (0.624268,0.362946)
 (0.624268,0.361894)
 (0.624268,0.360848)
 (0.624268,0.359809)
 (0.624268,0.358775)
 (0.624268,0.357747)
 (0.624268,0.356724)
 (0.624268,0.355708)
 (0.624268,0.354698)
 (0.624268,0.353693)
 (0.624268,0.352694)
 (0.624268,0.3517)
 (0.624268,0.350712)
 (0.624268,0.34973)
 (0.624268,0.348753)
 (0.624268,0.347782)
 (0.624268,0.346815)
 (0.624268,0.345855)
 (0.624268,0.344899)
 (0.624268,0.343949)
 (0.624268,0.343004)
 (0.624268,0.342065)
 (0.624268,0.34113)
 (0.624268,0.3402)
 (0.624268,0.339276)
 (0.624268,0.338357)
 (0.624268,0.337442)
 (0.624268,0.336533)
 (0.624268,0.335628)
 (0.624268,0.334728)
 (0.624268,0.333833)
 (0.624268,0.332943)
 (0.624268,0.332057)
 (0.624268,0.331177)
 (0.624268,0.3303)
 (0.624268,0.329429)
 (0.624268,0.328562)
 (0.624268,0.3277)
 (0.624268,0.326842)
 (0.624268,0.325988)
 (0.624268,0.325139)
 (0.624268,0.324295)
 (0.624268,0.323455)
 (0.624268,0.322619)
 (0.624268,0.321788)
 (0.624268,0.32096)
 (0.624268,0.320137)
 (0.624268,0.319319)
 (0.624268,0.318504)
 (0.624268,0.317694)
 (0.624268,0.316887)
 (0.624268,0.316085)
 (0.624268,0.315287)
 (0.624268,0.314493)
 (0.624268,0.313702)
 (0.624268,0.312916)
 (0.624268,0.312134)
 (0.624268,0.311356)
 (0.624268,0.310581)
 (0.624268,0.30981)
 (0.624268,0.309043)
 (0.624268,0.30828)
 (0.624268,0.307521)
 (0.624268,0.306766)
 (0.624268,0.306014)
 (0.624268,0.305265)
 (0.624268,0.304521)
 (0.624268,0.30378)
 (0.624268,0.303043)
 (0.624268,0.302309)
 (0.624268,0.301579)
 (0.624268,0.300852)
 (0.624268,0.300129)
 (0.624268,0.299409)
 (0.624268,0.298693)
 (0.624268,0.29798)
 (0.624268,0.29727)
 (0.624268,0.296564)
 (0.624268,0.295862)
 (0.624268,0.295162)
 (0.624268,0.294466)
 (0.624268,0.293773)
 (0.624268,0.293083)
 (0.624268,0.292397)
 (0.624268,0.291714)
 (0.624268,0.291034)
 (0.624268,0.290357)
 (0.624268,0.289683)
 (0.624268,0.289013)
 (0.624268,0.288345)
 (0.624268,0.287681)
 (0.624268,0.28702)
 (0.624268,0.286361)
 (0.624268,0.285706)
 (0.624268,0.285054)
 (0.624268,0.284404)
 (0.624268,0.283758)
 (0.624268,0.283115)
 (0.624268,0.282474)
 (0.624268,0.281836)
 (0.624268,0.281202)
 (0.624268,0.28057)
 (0.624268,0.279941)
 (0.624268,0.279314)
 (0.624268,0.278691)
 (0.624268,0.27807)
 (0.624268,0.277452)
 (0.624268,0.276837)
 (0.624268,0.276225)
 (0.624268,0.275615)
 (0.624268,0.275008)
 (0.624268,0.274403)
 (0.624268,0.273802)
 (0.624268,0.273203)
 (0.624268,0.272606)
 (0.624268,0.272012)
 (0.624268,0.271421)
 (0.624268,0.270832)
 (0.624268,0.270246)
 (0.624268,0.269662)
 (0.624268,0.269081)
 (0.624268,0.268502)
 (0.624268,0.267926)
 (0.624268,0.267352)
 (0.624268,0.266781)
 (0.624268,0.266212)
 (0.624268,0.265646)
 (0.624268,0.265082)
 (0.624268,0.26452)
 (0.624268,0.263961)
 (0.624268,0.263404)
 (0.624268,0.26285)
 (0.624268,0.262298)
 (0.624268,0.261748)
 (0.624268,0.2612)
 (0.624268,0.260655)
 (0.624268,0.260112)
 (0.624268,0.259571)
 (0.624268,0.259032)
 (0.624268,0.258496)
 (0.624268,0.257962)
 (0.624268,0.25743)
 (0.624268,0.2569)
 (0.624268,0.256373)
 (0.624268,0.255848)
 (0.624268,0.255324)
 (0.624268,0.254803)
 (0.624268,0.254284)
 (0.624268,0.253767)
 (0.624268,0.253253)
 (0.624268,0.25274)
 (0.624268,0.252229)
 (0.624268,0.251721)
 (0.624268,0.251214)
 (0.624268,0.25071)
 (0.624268,0.250207)
 (0.624268,0.249707)
 (0.624268,0.249209)
 (0.624268,0.248712)
 (0.624268,0.248218)
 (0.624268,0.247725)
 (0.624268,0.247235)
 (0.624268,0.246746)
 (0.624268,0.246259)
 (0.624268,0.245775)
 (0.624268,0.245292)
 (0.624268,0.244811)
 (0.624268,0.244332)
 (0.624268,0.243855)
 (0.624268,0.243379)
 (0.624268,0.242906)
 (0.624268,0.242434)
 (0.624268,0.241964)
 (0.624268,0.241496)
 (0.624268,0.24103)
 (0.624268,0.240566)
 (0.624268,0.240103)
 (0.624268,0.239642)
 (0.624268,0.239183)
 (0.624268,0.238726)
 (0.624268,0.23827)
 (0.624268,0.237816)
 (0.624268,0.237364)
 (0.624268,0.236914)
 (0.624268,0.236465)
 (0.624268,0.236018)
 (0.624268,0.235573)
 (0.624268,0.235129)
 (0.624268,0.234687)
 (0.624268,0.234247)
 (0.624268,0.233808)
 (0.624268,0.233371)
 (0.624268,0.232936)
 (0.624268,0.232502)
 (0.624268,0.23207)
 (0.624268,0.231639)
 (0.624268,0.23121)
 (0.624268,0.230783)
 (0.624268,0.230357)
 (0.624268,0.229933)
 (0.624268,0.22951)
 (0.624268,0.229089)
 (0.624268,0.228669)
 (0.624268,0.228251)
 (0.624268,0.227835)
 (0.624268,0.22742)
 (0.624268,0.227006)
 (0.624268,0.226594)
 (0.624268,0.226184)
 (0.624268,0.225775)
 (0.624268,0.225367)
 (0.624268,0.224961)
 (0.624268,0.224557)
 (0.624268,0.224154)
 (0.624268,0.223752)
 (0.624268,0.223352)
 (0.624268,0.222953)
 (0.624268,0.222555)
 (0.624268,0.222159)
 (0.624268,0.221765)
 (0.624268,0.221371)
 (0.624268,0.22098)
 (0.624268,0.220589)
 (0.624268,0.2202)
 (0.624268,0.219813)
 (0.624268,0.219426)
 (0.624268,0.219041)
 (0.624268,0.218658)
 (0.624268,0.218275)
 (0.624268,0.217894)
 (0.624268,0.217515)
 (0.624268,0.217137)
 (0.624268,0.21676)
 (0.624268,0.216384)
 (0.624268,0.21601)
 (0.624268,0.215636)
 (0.624268,0.215265)
 (0.624268,0.214894)
 (0.624268,0.214525)
 (0.624268,0.214157)
 (0.624268,0.21379)
 (0.624268,0.213425)
 (0.624268,0.213061)
 (0.624268,0.212698)
 (0.624268,0.212336)
 (0.624268,0.211975)
 (0.624268,0.211616)
 (0.624268,0.211258)
 (0.624268,0.210901)
 (0.624268,0.210546)
 (0.624268,0.210191)
 (0.624268,0.209838)
 (0.624268,0.209486)
 (0.624268,0.209135)
 (0.624268,0.208785)
 (0.624268,0.208437)
 (0.624268,0.208089)
 (0.624268,0)
} |- (axis cs:0,0) -- cycle;

\addplot [
color=red,
solid,
line width=1.5pt
]
coordinates{
 (0.75,0) (0.75,0.5) (0.5,0.75) (0,0.75)
};


\end{axis}

\end{tikzpicture}

%% file: puncturing.tex
In \cite{Barros-itsub03}, it is shown that correlated codes are suboptimal
when transmitting correlated sources over independent channels. The
conditions in (\ref{eq:sw}) implicitly assume the use of uncorrelated
codes i.e., we require the average mutual information (over the code
ensemble) $I(X_1;X_2) = 0$.

This condition is clearly not satisfied when we use a systematic LDPC
ensemble. This also explains the loss in performance of systematic
LDPC codes when compared to Turbo codes, as shown in
\cite{Martalo-ita10}. To ensure the independence of the transmitted
symbols, we use LDPC ensembles with punctured systematic encoders.


%% file: ldpc_de.tex
Assume that the sequences $\mathbf{U}_1$ and $\mathbf{U}_2$ are
encoded using LDPC codes with a degree distribution pair
$\left(\lambda,\rho\right)$ and a punctured systematic encoder. Let
the fraction of punctured (systematic) bits be $\gamma$.

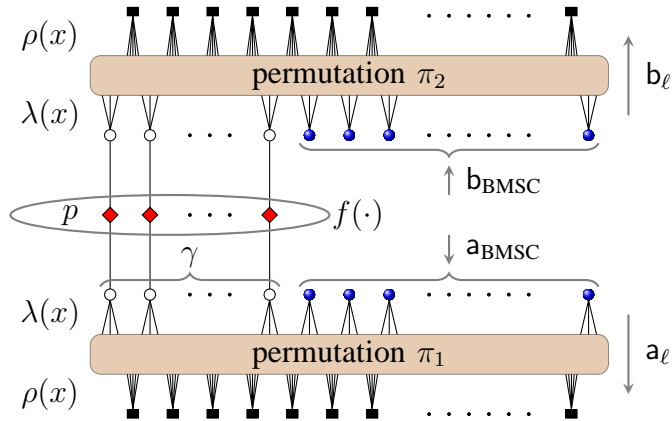
\begin{figure}
  \centering
  \input{tanner_ldpc.tex}
  \caption{Tanner Graph of an LDPC Code with source correlation}
  \label{fig:tanner_ldpc}
\end{figure}
The Tanner graph \cite{RU-2008} for the joint decoder is shown in
Figure~\ref{fig:tanner_ldpc}. Codes $1$ and $2$ correspond to the
bottom and top half of the graph. The codes are connected by
correlation nodes attached to the punctured bits. The joint iterative
decoder proceeds in rounds, by alternating one round of decoding for
code $1$ with one round of decoding for code $2$. Let $\a_{\ell}$ and
$\b_{\ell}$ denote the density\footnote{Assuming that the transmission
  alphabet is $\{\pm 1\}$, the densities are conditioned on the
  transmission of a $+1$.}  of the messages emanating from the
variable nodes at iteration $\ell$, corresponding to codes $1$ and
$2$. The density evolution equations \cite{RU-2008} can be written as
follows
\begin{equation}
  \label{eq:de-vc}
  \begin{split}
    \a_{\ell+1} &= \left[\gamma
      f\Bigl(L\left(\rho(\b_{\ell})\right)\Bigr) +
      (1-\gamma)\a_{\text{BMSC}}\right] \varoast \lambda(\rho(\a_{\ell})) \\
    \b_{\ell+1} &= \left[\gamma
      f\Bigl(L\left(\rho(\a_{\ell})\right)\Bigr) +
      (1-\gamma)\b_{\text{BMSC}}\right] \varoast \lambda(\rho(\b_{\ell})), \\
  \end{split}
\end{equation}
where $\lambda(\a)=\sum_i\lambda_i\a^{\varoast(i-1)}$,
$L(\a)=\sum_iL_i\a^{\varoast(i-1)}$,
$\rho(\a)=\sum_i\rho_i\a^{\boxast(i-1)}$, $\a_{\text{BMSC}}$ and
$\b_{\text{BMSC}}$ are the densities of the log-likelihood ratios received
from the channel. The function $f$ at the correlation nodes depends on
the equivalent channel corresponding to the correlation model, as
described in \cite{Chen-isit06}. Although one cannot assume that the
all-zero codeword is sent simultaneously by both users, one can show
that this DE recursion suffices for typical message pairs. 

First consider the BSC correlation model. By symmetry of the problem,
we can assume that user $1$ transmits the all-zero codeword and the
second user transmits a typical codeword. Due to the constraints
imposed by the correlation, the fraction of ones in the systematic
part of the codeword is $1-p$. Density evolution proceeds with two
types of messages (those connected to a variable node with transmitted
value $+1$ and those connected to a variable node with transmitted
value $-1$). By symmetry of the message passing rules
\cite[p. 210]{RU-2008}, we can factor out the sign for the messages
connected to variable nodes with transmitted value $-1$. This sign can
be factored into the correlation node (once again by the symmetry
condition). The fraction of correlation nodes which are flipped is
$1-p$. So, we introduce a parity-check at the correlation nodes which
evaluates to a Bernoulli-$p$ random variable i.e., $f(\a) =
\a_{\text{BSC}(p)}\boxast\a$. This simplification enables us to
proceed with density evolution assuming the transmission of an
all-zero codeword for both the users.

Note that such a simplification is not necessary for the erasure
correlation model. For a BEC correlation with probability $p$, there
is a parity-check at the correlation node with probability $p$ and
with probability $1-p$ there is no parity-check, so $f(\a) = (1-p) +
p\a$.

The residual error probability at iteration $\ell$,
$(e_1^{\ell},e_2^{\ell})$, is computed using the error functional
$\mathfrak{E}(\cdot)$ defined in \cite[p. 201]{RU-2008}:
\begin{align*}
  e_1^{\ell} &= \mathfrak{E}\left(\left[\gamma
      f\Bigl(L\left(\rho(\b_{\ell})\right)\Bigr) +
      (1-\gamma)\a_{\text{BMSC}}\right] \varoast L(\rho(\a_{\ell}))\right)\\
  e_2^{\ell} &= \mathfrak{E}\left(\left[\gamma
      f\Bigl(L\left(\rho(\a_{\ell})\right)\Bigr) +
      (1-\gamma)\b_{\text{BMSC}}\right] \varoast L(\rho(\b_{\ell}))\right).
\end{align*}


%% file: tanner_ldpc.tex
\begin{tikzpicture}[scale=0.53,>=stealth]

\draw[rounded corners,fill=brown,opacity=0.4] (-1,2) rectangle +(13,1);
\draw (5.5,2.5) node {permutation $\pi_1$};
\draw[rounded corners,fill=brown,opacity=0.4] (-1,9) rectangle +(13,1);
\draw (5.5,9.5) node {permutation $\pi_2$};

\foreach \x in {0,1,2,3,4,5,6,11}
{
  \filldraw[black] (\x,0.9)+(-2pt,0pt) rectangle +(6pt,6pt);
  \draw (\x,1)+(2pt,4pt) -- (\x,2);
  \draw (\x,1)+(2pt,4pt) -- ([xshift = 4pt]\x,2);
  \draw (\x,1)+(2pt,4pt) -- ([xshift = 2pt]\x,2);
  \draw (\x,1)+(2pt,4pt) -- ([xshift = -2pt]\x,2);
  \draw (\x,1)+(2pt,4pt) -- ([xshift = 6pt]\x,2);

  \filldraw[black] (\x,11)+(-2pt,0pt) rectangle +(6pt,6pt);
  \draw (\x,11)+(2pt,0pt) -- (\x,10);
  \draw (\x,11)+(2pt,0pt) -- ([xshift = 4pt]\x,10);
  \draw (\x,11)+(2pt,0pt) -- ([xshift = 2pt]\x,10);
  \draw (\x,11)+(2pt,0pt) -- ([xshift = -2pt]\x,10);
  \draw (\x,11)+(2pt,0pt) -- ([xshift = 6pt]\x,10);
}

\foreach \x in {-1,0,3} {
  \draw (\x,4)+(0.5,0) circle (4pt);
  \draw (\x,4)+(0.5cm,-4pt) -- ([xshift=0.5cm]\x,3);
  \draw (\x,4)+(0.5cm,-4pt) -- ([xshift=0.3cm]\x,3);
  \draw (\x,4)+(0.5cm,-4pt) -- ([xshift=0.7cm]\x,3);

  \draw (\x,8)+(0.5,0) circle (4pt);
  \draw (\x,8)+(0.5cm,4pt) -- ([xshift=0.5cm]\x,9);
  \draw (\x,8)+(0.5cm,4pt) -- ([xshift=0.3cm]\x,9);
  \draw (\x,8)+(0.5cm,4pt) -- ([xshift=0.7cm]\x,9);
}
\foreach \x in {4.5,5.5,6.5,11.5} {
  \shade[ball color=blue] (\x,4) circle (4pt);
  \draw (\x,4)+(0,-4pt) -- ([xshift=0cm]\x,3);
  \draw (\x,4)+(0,-4pt) -- ([xshift=-0.2cm]\x,3);
  \draw (\x,4)+(0,-4pt) -- ([xshift=0.2cm]\x,3);

  \shade[ball color=blue] (\x,8) circle (4pt);
  \draw (\x,8)+(0,4pt) -- ([xshift=0cm]\x,9);
  \draw (\x,8)+(0,4pt) -- ([xshift=-0.2cm]\x,9);
  \draw (\x,8)+(0,4pt) -- ([xshift=0.2cm]\x,9);
}

\foreach \x in {-0.5,0.5,3.5} {
  \node[diamond,draw=black,fill=red,inner sep=0pt,minimum size=6pt] at (\x,6) {};
}
\foreach \x in {1.5,2,2.5} {
  \foreach \y in {4,6,8} {
    \filldraw (\x,\y) circle (1pt);
  }
}
\begin{pgfonlayer}{background}
\draw (-1,4)+(0.5cm,4pt) -- ([xshift=0.5cm]-1,7.85);
\draw  (0,4)+(0.5cm,4pt) -- (0.5,7.85);
\draw  (3,4)+(0.5cm,4pt) -- (3.5,7.85);
\end{pgfonlayer}

\foreach \x in {7.5,8,8.5,...,10} {
  \foreach \y in {1,4,8,11} {
    \filldraw (\x,\y) circle (1pt);
  }
}

\draw (-2,1.5) node {$\rho(x)$};
\draw (-2,3.5) node {$\lambda(x)$};
\draw (-1.5,6) node {$p$};
\draw (-2,10.5) node {$\rho(x)$};
\draw (-2,8.5) node {$\lambda(x)$};

\draw [gray,thick,decorate,decoration={brace,amplitude=5pt}]
   (-0.75,4.2)  -- (3.75,4.2) 
   node [black,midway,above=3pt] {$\gamma$};

\draw[->,thick,gray] (8,5.5) -- (8,4.75) node[black,midway,right=2pt] {$\mathsf{a}_{\text{BMSC}}$};
\draw [gray,thick,decorate,decoration={brace,amplitude=5pt}]
   (4.25,4.2)  -- (11.75,4.2) ;
\draw[->,thick,gray] (8,6.5) -- (8,7.25) node[black,midway,right=2pt] {$\mathsf{b}_{\text{BMSC}}$};
\draw [gray,thick,decorate,decoration={brace,amplitude=5pt}]
    (11.75,7.8) -- (4.25,7.8) ;

\draw[gray,thick] (1,6) ellipse (4 cm and 0.5cm) node[black,right=2cm] {$f(\cdot)$};

\draw[->,thick,gray] (12.5,3.5) -- (12.5,1.5)  node[black,midway,right=2pt] {$\mathsf{a}_{\ell}$};
\draw[->,thick,gray] (12.5,8.5) -- (12.5,10.5) node[black,midway,right=2pt] {$\mathsf{b}_{\ell}$};
\end{tikzpicture}


%% file: staggering.tex
It is well known that single-user codes perform well at the corner
points of the SW region. Although single-user codes do not perform
well for symmetric channel conditions, they can be used to construct
staggered codes that perform well at the corner points and for
symmetric channel conditions. Consider $2$ sources with $Lk+(1 -
\beta)k$ bits each. Without loss of generality, add $\beta k$ zeros at
the beginning for source $U_1$ and add $\beta k$ zeros at the end for
source $U_2$, to get $(L+1)k$ bits. We call $\beta$ the staggering
fraction. Next encode each block of $k$ bits using a punctured
$(n-k,k)$ LDPC code. The rate loss incurred by the addition of $\beta
k$ zeros can be made arbitrarily small by increasing the number of
blocks $L$. At the decoder, one has the following structure:
\begin{figure}[t!]
  \centering
  \input{dec_structure_staggered}
  \caption{Decoder structure for staggered codes}
\end{figure}
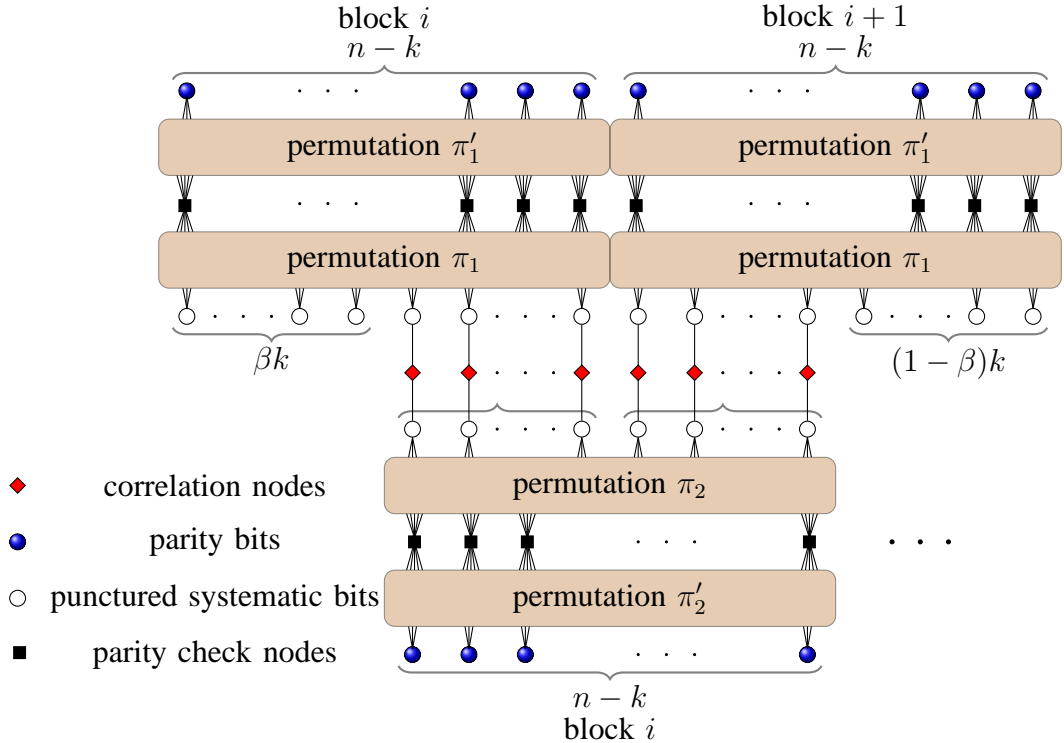
The performance of this staggered structure can be understood by
considering the erasure case in the limit $L\to\infty$.
\begin{theorem}
  Consider transmission over erasure channels with erasure rates
  $(\epsilon_1,\epsilon_2)$ using capacity approaching punctured
  $(n-k,k)$ LDPC codes. The staggered block code (with staggering
  fraction $\beta$) allows reliable communication for channel
  parameters
  \begin{align*}
    \epsilon_{1} &\leq \min\{1 - R(1- \beta),1 - R(1-p\beta)\},\text{
      and}\\\notag \epsilon_{2} &\leq 1 - R(1 - p(1-\beta)),
  \end{align*}
  where $R=k/(n-k)$ is the design rate of the code.
\end{theorem}
\begin{proof}
  Consider the first block for source $U_1$. The parity bits see a
  BEC$(\epsilon_1)$ channel and the source bits see an effective
  BEC$(1-\beta)$ channel (assuming no information comes from the
  decoder on the other side). So the effective erasure rate at the
  first block is $(1-R')\epsilon_{1} + R'(1-\beta)$ ($R'=k/n$ is the
  rate of the code before puncturing). The code can decode as long as
  $R' \leq 1 - ((1-R')\epsilon_{1} + R'(1-\beta))$ i.e., $\epsilon_{1}
  \leq 1 - R(1- \beta)$. Suppose the first block of $U_1$ can decode
  successfully, then the source bits in the first block of $U_2$ see
  an effective channel of $(1-\beta)(1-p) + \beta$. The parity bits
  see a channel with erasure probability $\epsilon_{2}$. So, the
  effective channel seen by the first block of the second code is
  $(1-R')\epsilon_{2} + R'(1 - p(1-\beta))$. So this block can be
  decoded as long as $\epsilon_{2} \leq 1 - R(1 - p(1-\beta))$. 
  The decoding continues by alternating between blocks of $U_1$ and
  $U_2$. This proves the claim.
\end{proof}
\begin{corollary}
  Consider transmission over erasure channels using capacity
  approaching punctured $(n-k,k)$ LDPC codes. The staggered block code
  (with staggering fraction $\beta=1/2$) allows reliable communication
  at both the corner points and the symmetric channel condition.
\end{corollary}
\begin{proof}
  The proof follows by matching the conditions of the previous theorem
  to a corner point and the extremal symmetric point of the SW
  region.  
\end{proof}
For general channels we can analyze the performance of the staggered
code using density evolution. Let $i\in\{1,\hdots,L\}$ and
$\a^{(i)}_{\ell}$ and $\b^{(i)}_{\ell}$ denote the density of the
messages emanating from the variable nodes at iteration $\ell$,
corresponding to codes $1$ and $2$ in block $i$. The DE equations can
be written as follows:
\begin{equation}
  \label{eq:de-vc-stag}
  \begin{split}
    \a^{(i)}_{\ell+1} &= \left[\gamma
      \left(\beta f\Bigl(L\left(\rho(\b^{(i-1)}_{\ell})\right)\Bigr) +
        (1-\beta) f\Bigl(L\left(\rho(\b^{(i)}_{\ell})\right)\Bigr)\right) +
      (1-\gamma)\a_{\text{BMSC}}\right] \varoast \lambda(\rho(\a_{\ell})) \\
    \b^{(i)}_{\ell+1} &= \left[\gamma
      \left((1-\beta)f\Bigl(L\left(\rho(\a^{(i)}_{\ell})\right)\Bigr)
        + \beta f\Bigl(L\left(\rho(\a^{(i+1)}_{\ell})\right)\Bigr)\right) +
      (1-\gamma)\b_{\text{BMSC}}\right] \varoast \lambda(\rho(\b_{\ell})).
  \end{split}
\end{equation}
Here, $\a^{(i)}_{\ell},\b^{(i)}_{\ell}=\Delta_{+\infty}$ (the delta
function at $\infty$) for $i\notin\{1,\hdots,L\}$.


%% file: dec_structure_staggered.tex
\begin{tikzpicture}[scale=0.75]
  \draw [gray,thick,decorate,decoration={brace,amplitude=5pt}]
   (0.25,1.2)  -- (3.75,1.2); 
  \draw [gray,thick,decorate,decoration={brace,amplitude=5pt}]
   (4.25,1.2)  -- (7.75,1.2) ;
  \foreach \x in {0.5,1.5,3.5,4.5,5.5,7.5} {
      \draw (\x,1) circle (4pt);
      \draw (\x,1.15) -- (\x,2.85);
      \node[diamond,draw=black,fill=red,inner sep=0pt,minimum size=6pt] at (\x,2) {};
      \draw (\x,0.85) -- ([xshift=-2pt]\x,0.5);
      \draw (\x,0.85) -- ([xshift= 0pt]\x,0.5);
      \draw (\x,0.85) -- ([xshift= 2pt]\x,0.5);
  }
  \foreach \x in {2,2.5,3,6,6.5,7} {
    \filldraw (\x,1) circle (0.5pt);
    \filldraw (\x,2) circle (0.5pt);
  }
  \foreach \x in {0.5,1.5,2.5,7.5} {
    \filldraw[black] (\x,-1)+(-2pt,-3pt) rectangle +(4pt,3pt);
    \draw (\x,-1)+(1pt,2pt) -- ([xshift = 1pt] \x,-0.5);
    \draw (\x,-1)+(1pt,2pt) -- ([xshift = 3pt] \x,-0.5);
    \draw (\x,-1)+(1pt,2pt) -- ([xshift = 5pt] \x,-0.5);
    \draw (\x,-1)+(1pt,2pt) -- ([xshift = -1pt]\x,-0.5);
    \draw (\x,-1)+(1pt,2pt) -- ([xshift = -3pt]\x,-0.5);
    \draw (\x,-1)+(1pt,0pt) -- ([xshift = 1pt] \x,-1.5);
    \draw (\x,-1)+(1pt,0pt) -- ([xshift = 3pt] \x,-1.5);
    \draw (\x,-1)+(1pt,0pt) -- ([xshift = 5pt] \x,-1.5);
    \draw (\x,-1)+(1pt,0pt) -- ([xshift = -1pt]\x,-1.5);
    \draw (\x,-1)+(1pt,0pt) -- ([xshift = -3pt]\x,-1.5);
    \draw[ball color=blue] (\x,-3) circle (4pt);
    \draw (\x,-2.9) -- ([xshift=-2pt]\x,-2.5);
    \draw (\x,-2.9) -- ([xshift= 0pt]\x,-2.5);
    \draw (\x,-2.9) -- ([xshift= 2pt]\x,-2.5);
  }
  \foreach \x in {4.5,5,5.5} {
    \filldraw (\x,-1) circle (0.5pt);
    \filldraw (\x,-3) circle (0.5pt);
  }
  \draw [gray,thick,decorate,decoration={brace,amplitude=5pt}]
   (7.75,-3.2)  -- (0.25,-3.2) node [black,midway,below=3pt] {$n- k$}
   node[black,midway,below=15pt] {block $i$};
   \draw[rounded corners,fill=brown,opacity=0.4] (0,-0.5) rectangle +(8,1);
   \draw (4,0) node {permutation $\pi_2$};
   \draw[rounded corners,fill=brown,opacity=0.4] (0,-2.5) rectangle +(8,1);
   \draw (4,-2) node {permutation $\pi_2'$};
   \begin{scope}[yshift=4cm,xshift=4cm,rotate=180]
  \draw [gray,thick,decorate,decoration={brace,amplitude=5pt}]
   (4.25,1.2)  -- (7.75,1.2) 
   node [black,midway,below=3pt] {$\beta k$};
  \foreach \x in {0.5,2.5,3.5,4.5,5.5,7.5} {
      \draw (\x,1) circle (4pt);
      \draw (\x,0.85) -- ([xshift=-2pt]\x,0.5);
      \draw (\x,0.85) -- ([xshift= 0pt]\x,0.5);
      \draw (\x,0.85) -- ([xshift= 2pt]\x,0.5);
  }
  \foreach \x in {1,1.5,2,6,6.5,7} {
    \filldraw (\x,1) circle (0.5pt);
  }
  \foreach \x in {0.5,1.5,2.5,7.5} {
    \filldraw[black] (\x,-1)+(-2pt,-2pt) rectangle +(4pt,4pt);
    \draw (\x,-1)+(1pt,2pt) -- ([xshift = 1pt] \x,-0.5);
    \draw (\x,-1)+(1pt,2pt) -- ([xshift = 3pt] \x,-0.5);
    \draw (\x,-1)+(1pt,2pt) -- ([xshift = 5pt] \x,-0.5);
    \draw (\x,-1)+(1pt,2pt) -- ([xshift = -1pt]\x,-0.5);
    \draw (\x,-1)+(1pt,2pt) -- ([xshift = -3pt]\x,-0.5);
    \draw (\x,-1)+(1pt,0pt) -- ([xshift = 1pt] \x,-1.5);
    \draw (\x,-1)+(1pt,0pt) -- ([xshift = 3pt] \x,-1.5);
    \draw (\x,-1)+(1pt,0pt) -- ([xshift = 5pt] \x,-1.5);
    \draw (\x,-1)+(1pt,0pt) -- ([xshift = -1pt]\x,-1.5);
    \draw (\x,-1)+(1pt,0pt) -- ([xshift = -3pt]\x,-1.5);
    \draw[ball color=blue] (\x,-3) circle (4pt);
    \draw (\x,-2.9) -- ([xshift=-2pt]\x,-2.5);
    \draw (\x,-2.9) -- ([xshift= 0pt]\x,-2.5);
    \draw (\x,-2.9) -- ([xshift= 2pt]\x,-2.5);
  }
  \foreach \x in {4.5,5,5.5} {
    \filldraw (\x,-1) circle (0.5pt);
    \filldraw (\x,-3) circle (0.5pt);
  }
  \draw [gray,thick,decorate,decoration={brace,amplitude=5pt}]
   (7.75,-3.2)  -- (0.25,-3.2) node [black,midway,above=3pt] {$n- k$}
   node[black,midway,above=15pt] {block $i$};
   \draw[rounded corners,fill=brown,opacity=0.4] (0,-0.5) rectangle +(8,1);
   \draw (4,0) node {permutation $\pi_1$};
   \draw[rounded corners,fill=brown,opacity=0.4] (0,-2.5) rectangle +(8,1);
   \draw (4,-2) node {permutation $\pi_1'$};
   \end{scope}
   \begin{scope}[yshift=4cm,xshift=12cm,rotate=180]
       \draw [gray,thick,decorate,decoration={brace,amplitude=5pt}]
   (0.25,1.2)  -- (3.75,1.2) 
   node [black,midway,below=3pt] {$(1-\beta) k$};
  \foreach \x in {0.5,1.5,3.5,4.5,6.5,7.5} {
      \draw (\x,1) circle (4pt);
      \draw (\x,0.85) -- ([xshift=-2pt]\x,0.5);
      \draw (\x,0.85) -- ([xshift= 0pt]\x,0.5);
      \draw (\x,0.85) -- ([xshift= 2pt]\x,0.5);
  }
  \foreach \x in {2,2.5,3,5,5.5,6} {
    \filldraw (\x,1) circle (0.5pt);
  }
  \foreach \x in {0.5,1.5,2.5,7.5} {
    \filldraw[black] (\x,-1)+(-2pt,-2pt) rectangle +(4pt,4pt);
    \draw (\x,-1)+(1pt,2pt) -- ([xshift = 1pt] \x,-0.5);
    \draw (\x,-1)+(1pt,2pt) -- ([xshift = 3pt] \x,-0.5);
    \draw (\x,-1)+(1pt,2pt) -- ([xshift = 5pt] \x,-0.5);
    \draw (\x,-1)+(1pt,2pt) -- ([xshift = -1pt]\x,-0.5);
    \draw (\x,-1)+(1pt,2pt) -- ([xshift = -3pt]\x,-0.5);
    \draw (\x,-1)+(1pt,0pt) -- ([xshift = 1pt] \x,-1.5);
    \draw (\x,-1)+(1pt,0pt) -- ([xshift = 3pt] \x,-1.5);
    \draw (\x,-1)+(1pt,0pt) -- ([xshift = 5pt] \x,-1.5);
    \draw (\x,-1)+(1pt,0pt) -- ([xshift = -1pt]\x,-1.5);
    \draw (\x,-1)+(1pt,0pt) -- ([xshift = -3pt]\x,-1.5);
    \draw[ball color=blue] (\x,-3) circle (4pt);
    \draw (\x,-2.9) -- ([xshift=-2pt]\x,-2.5);
    \draw (\x,-2.9) -- ([xshift= 0pt]\x,-2.5);
    \draw (\x,-2.9) -- ([xshift= 2pt]\x,-2.5);
  }
  \foreach \x in {4.5,5,5.5} {
    \filldraw (\x,-1) circle (0.5pt);
    \filldraw (\x,-3) circle (0.5pt);
  }
  \draw [gray,thick,decorate,decoration={brace,amplitude=5pt}]
   (7.75,-3.2)  -- (0.25,-3.2) node [black,midway,above=3pt] {$n- k$}
   node[black,midway,above=15pt] {block $i+1$};
   \draw[rounded corners,fill=brown,opacity=0.4] (0,-0.5) rectangle +(8,1);
   \draw (4,0) node {permutation $\pi_1$};
   \draw[rounded corners,fill=brown,opacity=0.4] (0,-2.5) rectangle +(8,1);
   \draw (4,-2) node {permutation $\pi_1'$};
   \end{scope}
   \foreach \x in {9,9.5,10} {
     \filldraw (\x,-1) circle (1pt);
   }
   \node[diamond,draw=black,fill=red,inner sep=0pt,minimum size=6pt] at (-6.5,0) {};
    \node (leg4) at (-3,0) {correlation nodes};
    \draw[ball color=blue] (-6.5,-1) circle (4pt);
    \node (leg1) at (-3,-1) {parity bits};
    \draw (-6.5,-2) circle (4pt);
    \node (leg2) at (-3,-2) {punctured systematic bits};
    \filldraw[black] (-6.5,-3)+(-2pt,-2pt) rectangle +(4pt,4pt);
    \node (leg3) at (-3,-3) {parity check nodes};
   
\end{tikzpicture}

%% file: diff_ev.tex
Throughout this section, we use $x$ to denote an element of
$\mathbb{R}^n$ for some $n\in\mathbb{N}$, and $x_i$ to denote its
$i$th component.  Let $\mathcal{V} = \{i\,|\,\lambda_i\neq 0\}$ and
$\mathcal{P} = \{i\,|\,\rho_i\neq 0\}$ be the support sets of the
variable and parity-check degree distributions respectively, which are
assumed to be known. The correlation parameter $p$ is fixed. We design
LDPC codes for this scenario using differential evolution
\cite{Price-2005}, for a design rate $R_d$.
Let
\begin{align*}
  \Delta^{n-1} = \left\{ x\in \mathbb{R}^{n} \middle |
    \sum_{i=1}^{n}x_{i}=1,x_{i}\geq0,i=1,\cdots,n \right\}
\end{align*}
denote the unit simplex and $n_v=\abs{\mathcal{V}}$,
$n_p=\abs{\mathcal{P}}$. Then, the search space for all variable
(check) degree profiles is $\Delta^{n_v-1}$ ($\Delta^{n_p-1}$).  The
optimization is performed over the search space $\mathcal{S} =
\Delta^{n_v-1}\times \Delta^{n_p-1}$, with parameter vectors $x =
[x_{\lambda}, x_{\rho}]$\footnote{$(x_{\lambda},\mathcal{V})$ and
  $(x_{\rho},\mathcal{P})$ correspond to the variable and parity node
  degree profiles respectively.}, where $x_{\lambda} \in
\Delta^{n_v-1}, x_{\rho} \in \Delta^{n_p-1}$. In our optimization
procedure, we expand the search space to $\mathcal{S}' =
\{x\in\mathbb{R}^{n_v+n_p}, \sum_i(x_{\lambda})_i=1,
\sum_i(x_{\rho})_i=1\}$, for simplicity in the crossover stage. We
generate an initial population of trial degree distributions by
uniformly sampling the degree distributions from the unit simplex.


Let $\mathsf{C}$ be a finite subset of channel parameters
$(\alpha_1,\alpha_2)$ that correspond to the sum rate constraint of
the \sw{} conditions for a design rate $R_d$. Let $\Gamma:
\mathcal{S}' \times \mathsf{C} \to [0,1]\times[0,1]$,
$(x,\alpha_1,\alpha_2)\mapsto (e_1,e_2)$ be the function that gives
the residual error probability\footnote{We set the maximum number of
  iterations to $100$ for all the designs considered in this
  paper. Density evolution is stopped when the maximum number of
  iterations is reached or the difference in the residual error
  probability between successive iterations is less than $10^{-8}$.}
(using joint density evolution as described in
Section~\ref{sec:density-evolution}) for each decoder, for a pair of
codes with degree distribution $x$ $\left(\text{i.e.,
  }(x_{\lambda},x_{\rho})\right)$, when transmitted over channels with
parameters $(\alpha_1,\alpha_2)$. We use discretized density evolution
\cite{Chung-comlett01}\footnote{A $9$ bit linear quantization is used
  over a likelihood ratio range $[-20, 20]$} to compute the
performance of an ensemble.

For our design, we want the code to achieve an arbitrarily low
probability of error on $\mathsf{C}$ and we want the rate of the code
$R(x)$ to be as close to the design rate $R_d$ as possible. So, we
define the cost function,
\begin{equation*}
  \begin{split}
    \mathcal{F}(x) &= a \cdot \left(
      \sum_{(\alpha_1,\alpha_2)\in\mathcal{C}} \left(1-
        \indicator{(\alpha_1,\alpha_2) |
          \Gamma(x,\alpha_1,\alpha_2)\preceq(\tau,\tau)}\right)\right) + b\cdot(R_d-R(x)),
  \end{split}
\end{equation*}
if $x\in\mathcal{S}$ and $\mathcal{F}(x)=\infty$, if
$x\in\mathcal{S}'\backslash\mathcal{S}$. 
The constants $a$ and $b$ are chosen
through trial and error. The parameters chosen for the designs
considered in this paper are $\tau=10^{-5}, a = 10$ and $b = 30$. The
optimization is then setup as $\min_{x\in\mathcal{S}'}\mathcal{F}(x)$.

We use a variant of differential evolution, with the mutation
and recombination scheme given in \cite{Shokrollahi-isit00}. 
The resulting codes are then staggered as described in
Section~\ref{sec:single-user-codes}.

%% file: results.tex
This paper shows that the \sw{} conditions are necessary and sufficient for
communication of correlated sources through independent BMS channels,
without channel state information at the transmitter. This implies
that a single random code is sufficient to communicate with vanishing
probability of error, for the entire \sw{} region. We showed the
achievability of the symmetric channel condition under message passing
by providing a sequence of LDGM ensembles which can achieve an
arbitrarly low probability of error.

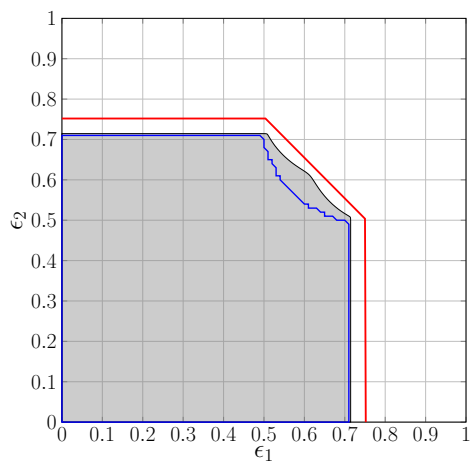
\begin{figure}[b!]
  \centering
  \input{erasure_opt_roc}
  \vspace{-3mm}
  \caption{ACPR (Density Evolution threshold) of an optimized (erasure
    channel) LDPC Code of rate $0.3308$ is shown in blue. The grey
    area is the ACPR after staggering.}
  \label{fig:roc_ldpc_erasure}
\end{figure}

We designed punctured systematic LDPC codes for the scenarios
described in Section~\ref{sec:problem-setup}. The design was performed
to maximize the ACPR, in contrast to previous work. For the erasure
correlation model, the optimization
was performed for a design rate of $R_d=0.57$ after puncturing and
source correlation $p=0.5$. The resulting degree profile
\begin{align*}
  \lambda(x) &= 0.3633 x + 0.2834 x^{2} + 0.2315 x^{6} + 0.1217 x^{19}, \\
  \rho(x) &= 0.531776 x^{3} + 0.468224 x^{5},
\end{align*}
has a design rate of $0.3308$ and transmission rate $0.4962$. The ACPR
for this code is shown in Figure~\ref{fig:roc_ldpc_erasure} along with
the \sw{} region for the rate pair $(0.4962,0.4962)$.  This shows
optimized ensembles can achieve a large portion of the \sw{} region.

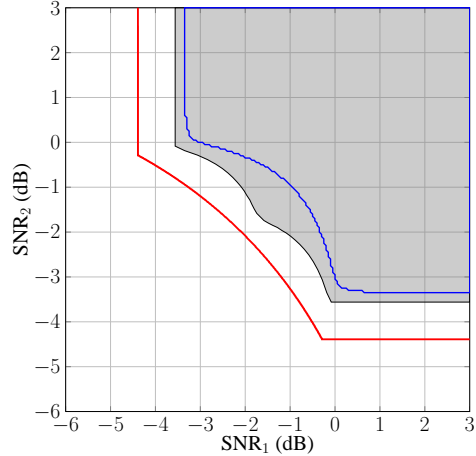
\begin{figure}[t!]
  \centering
  \input{awgn_opt_roc}
  \vspace{-3mm}
  \caption{ACPR (Density Evolution threshold) of an optimized (AWGN
    channel) LDPC Code of rate $0.323$ is shown in blue. The grey area
  is the ACPR after staggering.}
  \label{fig:roc_ldpc_awgn}
\end{figure}

The BSC source correlation parameter was $p=0.9$ and the optimization
was performed for a design rate $R_d=0.5$ after puncturing. The resulting
degree profile
\begin{align*}
  \lambda(x) &= 0.26725x + 0.26823x^{2} + 0.07557x^{3} + 0.212x^{6} + 0.027898x^{7} + 0.0061593x^{8}+\\
 &\phantom{==}0.0011654x^{14} +0.14173x^{19},\\
  \rho(x) &= 0.37856x^{3} + 0.56211x^{5} + 0.0080803x^{9}
  +0.028448x^{14} + 0.0095319x^{19} +0.013267x^{24},
\end{align*}
has a design rate of $0.323$ and transmission rate $0.476$. The ACPR
for this code is shown in Figure~\ref{fig:roc_ldpc_awgn} along with
the \sw{} region for the rate pair $(0.476,0.476)$. These results show
that ensembles optimized using differential evolution almost achieve
the entire \sw{} region. 

%% file: erasure_opt_roc.tex
%

\begin{tikzpicture}[scale=0.47]

\begin{axis}[
scale only axis,
width=4.5in,
height=4.5in,
xmin=0, xmax=1,
ymin=0, ymax=1,
xtick={0,0.1,0.2,0.3,0.4,0.5,0.6,0.7,0.8,0.9,1},
ytick={0,0.1,0.2,0.3,0.4,0.5,0.6,0.7,0.8,0.9,1},
xlabel={\LARGE$\epsilon_1$},
ylabel={\LARGE$\epsilon_2$},
xmajorgrids,
ymajorgrids]

\addplot [
color=black,fill=gray,fill opacity=0.4]
coordinates{
  (0,0.714478)
 (0.238159,0.714478)
 (0.238557,0.714478)
 (0.238956,0.714478)
 (0.239356,0.714478)
 (0.239758,0.714478)
 (0.240161,0.714478)
 (0.240565,0.714478)
 (0.240971,0.714478)
 (0.241378,0.714478)
 (0.241786,0.714478)
 (0.242196,0.714478)
 (0.242607,0.714478)
 (0.24302,0.714478)
 (0.243434,0.714478)
 (0.243849,0.714478)
 (0.244266,0.714478)
 (0.244684,0.714478)
 (0.245104,0.714478)
 (0.245525,0.714478)
 (0.245948,0.714478)
 (0.246372,0.714478)
 (0.246797,0.714478)
 (0.247224,0.714478)
 (0.247653,0.714478)
 (0.248082,0.714478)
 (0.248514,0.714478)
 (0.248947,0.714478)
 (0.249381,0.714478)
 (0.249817,0.714478)
 (0.250255,0.714478)
 (0.250694,0.714478)
 (0.251134,0.714478)
 (0.251577,0.714478)
 (0.25202,0.714478)
 (0.252466,0.714478)
 (0.252912,0.714478)
 (0.253361,0.714478)
 (0.253811,0.714478)
 (0.254262,0.714478)
 (0.254716,0.714478)
 (0.255171,0.714478)
 (0.255627,0.714478)
 (0.256085,0.714478)
 (0.256545,0.714478)
 (0.257006,0.714478)
 (0.257469,0.714478)
 (0.257934,0.714478)
 (0.258401,0.714478)
 (0.258869,0.714478)
 (0.259338,0.714478)
 (0.25981,0.714478)
 (0.260283,0.714478)
 (0.260758,0.714478)
 (0.261235,0.714478)
 (0.261713,0.714478)
 (0.262194,0.714478)
 (0.262676,0.714478)
 (0.263159,0.714478)
 (0.263645,0.714478)
 (0.264132,0.714478)
 (0.264621,0.714478)
 (0.265112,0.714478)
 (0.265605,0.714478)
 (0.2661,0.714478)
 (0.266596,0.714478)
 (0.267094,0.714478)
 (0.267595,0.714478)
 (0.268097,0.714478)
 (0.268601,0.714478)
 (0.269106,0.714478)
 (0.269614,0.714478)
 (0.270124,0.714478)
 (0.270635,0.714478)
 (0.271149,0.714478)
 (0.271664,0.714478)
 (0.272182,0.714478)
 (0.272701,0.714478)
 (0.273223,0.714478)
 (0.273746,0.714478)
 (0.274272,0.714478)
 (0.274799,0.714478)
 (0.275329,0.714478)
 (0.27586,0.714478)
 (0.276394,0.714478)
 (0.276929,0.714478)
 (0.277467,0.714478)
 (0.278007,0.714478)
 (0.278549,0.714478)
 (0.279093,0.714478)
 (0.279639,0.714478)
 (0.280187,0.714478)
 (0.280738,0.714478)
 (0.28129,0.714478)
 (0.281845,0.714478)
 (0.282402,0.714478)
 (0.282961,0.714478)
 (0.283523,0.714478)
 (0.284086,0.714478)
 (0.284652,0.714478)
 (0.285221,0.714478)
 (0.285791,0.714478)
 (0.286364,0.714478)
 (0.286939,0.714478)
 (0.287516,0.714478)
 (0.288096,0.714478)
 (0.288678,0.714478)
 (0.289262,0.714478)
 (0.289849,0.714478)
 (0.290438,0.714478)
 (0.29103,0.714478)
 (0.291623,0.714478)
 (0.29222,0.714478)
 (0.292819,0.714478)
 (0.29342,0.714478)
 (0.294024,0.714478)
 (0.29463,0.714478)
 (0.295239,0.714478)
 (0.29585,0.714478)
 (0.296464,0.714478)
 (0.29708,0.714478)
 (0.297699,0.714478)
 (0.29832,0.714478)
 (0.298945,0.714478)
 (0.299571,0.714478)
 (0.300201,0.714478)
 (0.300833,0.714478)
 (0.301467,0.714478)
 (0.302105,0.714478)
 (0.302745,0.714478)
 (0.303387,0.714478)
 (0.304033,0.714478)
 (0.304681,0.714478)
 (0.305332,0.714478)
 (0.305986,0.714478)
 (0.306643,0.714478)
 (0.307302,0.714478)
 (0.307964,0.714478)
 (0.30863,0.714478)
 (0.309298,0.714478)
 (0.309969,0.714478)
 (0.310642,0.714478)
 (0.311319,0.714478)
 (0.311999,0.714478)
 (0.312682,0.714478)
 (0.313367,0.714478)
 (0.314056,0.714478)
 (0.314748,0.714478)
 (0.315443,0.714478)
 (0.316141,0.714478)
 (0.316841,0.714478)
 (0.317546,0.714478)
 (0.318253,0.714478)
 (0.318963,0.714478)
 (0.319677,0.714478)
 (0.320394,0.714478)
 (0.321114,0.714478)
 (0.321837,0.714478)
 (0.322563,0.714478)
 (0.323293,0.714478)
 (0.324026,0.714478)
 (0.324763,0.714478)
 (0.325502,0.714478)
 (0.326245,0.714478)
 (0.326992,0.714478)
 (0.327742,0.714478)
 (0.328495,0.714478)
 (0.329252,0.714478)
 (0.330013,0.714478)
 (0.330777,0.714478)
 (0.331544,0.714478)
 (0.332315,0.714478)
 (0.33309,0.714478)
 (0.333868,0.714478)
 (0.33465,0.714478)
 (0.335435,0.714478)
 (0.336225,0.714478)
 (0.337018,0.714478)
 (0.337814,0.714478)
 (0.338615,0.714478)
 (0.339419,0.714478)
 (0.340227,0.714478)
 (0.341039,0.714478)
 (0.341855,0.714478)
 (0.342675,0.714478)
 (0.343499,0.714478)
 (0.344327,0.714478)
 (0.345158,0.714478)
 (0.345994,0.714478)
 (0.346834,0.714478)
 (0.347678,0.714478)
 (0.348526,0.714478)
 (0.349378,0.714478)
 (0.350234,0.714478)
 (0.351095,0.714478)
 (0.351959,0.714478)
 (0.352828,0.714478)
 (0.353702,0.714478)
 (0.354579,0.714478)
 (0.355461,0.714478)
 (0.356348,0.714478)
 (0.357239,0.714478)
 (0.358134,0.714478)
 (0.359034,0.714478)
 (0.359938,0.714478)
 (0.360847,0.714478)
 (0.361761,0.714478)
 (0.362679,0.714478)
 (0.363602,0.714478)
 (0.364529,0.714478)
 (0.365462,0.714478)
 (0.366399,0.714478)
 (0.367341,0.714478)
 (0.368287,0.714478)
 (0.369239,0.714478)
 (0.370196,0.714478)
 (0.371157,0.714478)
 (0.372124,0.714478)
 (0.373095,0.714478)
 (0.374072,0.714478)
 (0.375054,0.714478)
 (0.376041,0.714478)
 (0.377033,0.714478)
 (0.37803,0.714478)
 (0.379033,0.714478)
 (0.380041,0.714478)
 (0.381055,0.714478)
 (0.382074,0.714478)
 (0.383098,0.714478)
 (0.384128,0.714478)
 (0.385163,0.714478)
 (0.386204,0.714478)
 (0.387251,0.714478)
 (0.388303,0.714478)
 (0.389361,0.714478)
 (0.390425,0.714478)
 (0.391495,0.714478)
 (0.39257,0.714478)
 (0.393652,0.714478)
 (0.394739,0.714478)
 (0.395832,0.714478)
 (0.396932,0.714478)
 (0.398038,0.714478)
 (0.399149,0.714478)
 (0.400268,0.714478)
 (0.401392,0.714478)
 (0.402523,0.714478)
 (0.40366,0.714478)
 (0.404803,0.714478)
 (0.405953,0.714478)
 (0.40711,0.714478)
 (0.408273,0.714478)
 (0.409443,0.714478)
 (0.410619,0.714478)
 (0.411803,0.714478)
 (0.412993,0.714478)
 (0.41419,0.714478)
 (0.415394,0.714478)
 (0.416605,0.714478)
 (0.417823,0.714478)
 (0.419048,0.714478)
 (0.420281,0.714478)
 (0.421521,0.714478)
 (0.422768,0.714478)
 (0.424022,0.714478)
 (0.425284,0.714478)
 (0.426554,0.714478)
 (0.427831,0.714478)
 (0.429116,0.714478)
 (0.430408,0.714478)
 (0.431708,0.714478)
 (0.433017,0.714478)
 (0.434333,0.714478)
 (0.435657,0.714478)
 (0.436989,0.714478)
 (0.43833,0.714478)
 (0.439678,0.714478)
 (0.441036,0.714478)
 (0.442401,0.714478)
 (0.443775,0.714478)
 (0.445157,0.714478)
 (0.446548,0.714478)
 (0.447948,0.714478)
 (0.449357,0.714478)
 (0.450774,0.714478)
 (0.452201,0.714478)
 (0.453637,0.714478)
 (0.455081,0.714478)
 (0.456535,0.714478)
 (0.457998,0.714478)
 (0.459471,0.714478)
 (0.460953,0.714478)
 (0.462445,0.714478)
 (0.463946,0.714478)
 (0.465458,0.714478)
 (0.466979,0.714478)
 (0.46851,0.714478)
 (0.470051,0.714478)
 (0.471602,0.714478)
 (0.473164,0.714478)
 (0.474736,0.714478)
 (0.476318,0.714478)
 (0.477911,0.714478)
 (0.479515,0.714478)
 (0.48113,0.714478)
 (0.482755,0.714478)
 (0.484392,0.714478)
 (0.486039,0.714478)
 (0.487698,0.714478)
 (0.489368,0.714478)
 (0.49105,0.714478)
 (0.492743,0.714478)
 (0.494448,0.714478)
 (0.496165,0.714478)
 (0.497894,0.714478)
 (0.499592,0.714417)
 (0.501345,0.714417)
 (0.503089,0.714386)
 (0.504759,0.714233)
 (0.506375,0.713989)
 (0.507765,0.713409)
 (0.508772,0.71228)
 (0.50972,0.71106)
 (0.5105,0.709595)
 (0.511417,0.708313)
 (0.512253,0.706909)
 (0.51305,0.705444)
 (0.513954,0.704117)
 (0.514797,0.702698)
 (0.515657,0.701294)
 (0.516479,0.699829)
 (0.51742,0.698517)
 (0.518277,0.697083)
 (0.519163,0.695679)
 (0.520125,0.694366)
 (0.521047,0.692993)
 (0.522002,0.691652)
 (0.522914,0.690247)
 (0.523835,0.688843)
 (0.524856,0.687561)
 (0.525791,0.686157)
 (0.526827,0.684875)
 (0.527777,0.683472)
 (0.528877,0.682251)
 (0.529913,0.680939)
 (0.53091,0.679565)
 (0.531987,0.678284)
 (0.533084,0.677017)
 (0.534118,0.675659)
 (0.535269,0.674438)
 (0.53638,0.673157)
 (0.537503,0.671879)
 (0.53858,0.670532)
 (0.539767,0.669312)
 (0.540915,0.66803)
 (0.542121,0.666809)
 (0.543313,0.665558)
 (0.544526,0.664322)
 (0.5457,0.663025)
 (0.546946,0.661804)
 (0.548202,0.660583)
 (0.549469,0.659363)
 (0.550747,0.658142)
 (0.552035,0.656921)
 (0.553334,0.655701)
 (0.554697,0.654543)
 (0.55603,0.653336)
 (0.557376,0.65213)
 (0.558747,0.65094)
 (0.560103,0.649719)
 (0.56155,0.64859)
 (0.562956,0.6474)
 (0.564415,0.646255)
 (0.56586,0.645081)
 (0.567345,0.643936)
 (0.568815,0.642761)
 (0.57034,0.641632)
 (0.571823,0.640442)
 (0.573402,0.639343)
 (0.574968,0.638214)
 (0.576551,0.637089)
 (0.578114,0.635925)
 (0.57975,0.634827)
 (0.581374,0.633698)
 (0.58304,0.632599)
 (0.584708,0.631485)
 (0.586393,0.630373)
 (0.588048,0.629211)
 (0.589777,0.628113)
 (0.591523,0.627014)
 (0.593256,0.625885)
 (0.59502,0.624771)
 (0.596743,0.623596)
 (0.598526,0.622467)
 (0.600267,0.621277)
 (0.60207,0.620132)
 (0.603816,0.618912)
 (0.605505,0.617615)
 (0.607255,0.616364)
 (0.608916,0.615005)
 (0.610534,0.613586)
 (0.612091,0.612091)
 (0.612091,0.612091)
 (0.613586,0.610534)
 (0.615005,0.608916)
 (0.616364,0.607255)
 (0.617615,0.605505)
 (0.618912,0.603816)
 (0.620132,0.60207)
 (0.621277,0.600267)
 (0.622467,0.598526)
 (0.623596,0.596743)
 (0.624771,0.59502)
 (0.625885,0.593256)
 (0.627014,0.591523)
 (0.628113,0.589777)
 (0.629211,0.588048)
 (0.630373,0.586393)
 (0.631485,0.584708)
 (0.632599,0.58304)
 (0.633698,0.581374)
 (0.634827,0.57975)
 (0.635925,0.578114)
 (0.637089,0.576551)
 (0.638214,0.574968)
 (0.639343,0.573402)
 (0.640442,0.571823)
 (0.641632,0.57034)
 (0.642761,0.568815)
 (0.643936,0.567345)
 (0.645081,0.56586)
 (0.646255,0.564415)
 (0.6474,0.562956)
 (0.64859,0.56155)
 (0.649719,0.560103)
 (0.65094,0.558747)
 (0.65213,0.557376)
 (0.653336,0.55603)
 (0.654543,0.554697)
 (0.655701,0.553334)
 (0.656921,0.552035)
 (0.658142,0.550747)
 (0.659363,0.549469)
 (0.660583,0.548202)
 (0.661804,0.546946)
 (0.663025,0.5457)
 (0.664322,0.544526)
 (0.665558,0.543313)
 (0.666809,0.542121)
 (0.66803,0.540915)
 (0.669312,0.539767)
 (0.670532,0.53858)
 (0.671879,0.537503)
 (0.673157,0.53638)
 (0.674438,0.535269)
 (0.675659,0.534118)
 (0.677017,0.533084)
 (0.678284,0.531987)
 (0.679565,0.53091)
 (0.680939,0.529913)
 (0.682251,0.528877)
 (0.683472,0.527777)
 (0.684875,0.526827)
 (0.686157,0.525791)
 (0.687561,0.524856)
 (0.688843,0.523835)
 (0.690247,0.522914)
 (0.691652,0.522002)
 (0.692993,0.521047)
 (0.694366,0.520125)
 (0.695679,0.519163)
 (0.697083,0.518277)
 (0.698517,0.51742)
 (0.699829,0.516479)
 (0.701294,0.515657)
 (0.702698,0.514797)
 (0.704117,0.513954)
 (0.705444,0.51305)
 (0.706909,0.512253)
 (0.708313,0.511417)
 (0.709595,0.5105)
 (0.71106,0.50972)
 (0.71228,0.508772)
 (0.713409,0.507765)
 (0.713989,0.506375)
 (0.714233,0.504759)
 (0.714386,0.503089)
 (0.714417,0.501345)
 (0.714417,0.499592)
 (0.714478,0.497894)
 (0.714478,0.496165)
 (0.714478,0.494448)
 (0.714478,0.492743)
 (0.714478,0.49105)
 (0.714478,0.489368)
 (0.714478,0.487698)
 (0.714478,0.486039)
 (0.714478,0.484392)
 (0.714478,0.482755)
 (0.714478,0.48113)
 (0.714478,0.479515)
 (0.714478,0.477911)
 (0.714478,0.476318)
 (0.714478,0.474736)
 (0.714478,0.473164)
 (0.714478,0.471602)
 (0.714478,0.470051)
 (0.714478,0.46851)
 (0.714478,0.466979)
 (0.714478,0.465458)
 (0.714478,0.463946)
 (0.714478,0.462445)
 (0.714478,0.460953)
 (0.714478,0.459471)
 (0.714478,0.457998)
 (0.714478,0.456535)
 (0.714478,0.455081)
 (0.714478,0.453637)
 (0.714478,0.452201)
 (0.714478,0.450774)
 (0.714478,0.449357)
 (0.714478,0.447948)
 (0.714478,0.446548)
 (0.714478,0.445157)
 (0.714478,0.443775)
 (0.714478,0.442401)
 (0.714478,0.441036)
 (0.714478,0.439678)
 (0.714478,0.43833)
 (0.714478,0.436989)
 (0.714478,0.435657)
 (0.714478,0.434333)
 (0.714478,0.433017)
 (0.714478,0.431708)
 (0.714478,0.430408)
 (0.714478,0.429116)
 (0.714478,0.427831)
 (0.714478,0.426554)
 (0.714478,0.425284)
 (0.714478,0.424022)
 (0.714478,0.422768)
 (0.714478,0.421521)
 (0.714478,0.420281)
 (0.714478,0.419048)
 (0.714478,0.417823)
 (0.714478,0.416605)
 (0.714478,0.415394)
 (0.714478,0.41419)
 (0.714478,0.412993)
 (0.714478,0.411803)
 (0.714478,0.410619)
 (0.714478,0.409443)
 (0.714478,0.408273)
 (0.714478,0.40711)
 (0.714478,0.405953)
 (0.714478,0.404803)
 (0.714478,0.40366)
 (0.714478,0.402523)
 (0.714478,0.401392)
 (0.714478,0.400268)
 (0.714478,0.399149)
 (0.714478,0.398038)
 (0.714478,0.396932)
 (0.714478,0.395832)
 (0.714478,0.394739)
 (0.714478,0.393652)
 (0.714478,0.39257)
 (0.714478,0.391495)
 (0.714478,0.390425)
 (0.714478,0.389361)
 (0.714478,0.388303)
 (0.714478,0.387251)
 (0.714478,0.386204)
 (0.714478,0.385163)
 (0.714478,0.384128)
 (0.714478,0.383098)
 (0.714478,0.382074)
 (0.714478,0.381055)
 (0.714478,0.380041)
 (0.714478,0.379033)
 (0.714478,0.37803)
 (0.714478,0.377033)
 (0.714478,0.376041)
 (0.714478,0.375054)
 (0.714478,0.374072)
 (0.714478,0.373095)
 (0.714478,0.372124)
 (0.714478,0.371157)
 (0.714478,0.370196)
 (0.714478,0.369239)
 (0.714478,0.368287)
 (0.714478,0.367341)
 (0.714478,0.366399)
 (0.714478,0.365462)
 (0.714478,0.364529)
 (0.714478,0.363602)
 (0.714478,0.362679)
 (0.714478,0.361761)
 (0.714478,0.360847)
 (0.714478,0.359938)
 (0.714478,0.359034)
 (0.714478,0.358134)
 (0.714478,0.357239)
 (0.714478,0.356348)
 (0.714478,0.355461)
 (0.714478,0.354579)
 (0.714478,0.353702)
 (0.714478,0.352828)
 (0.714478,0.351959)
 (0.714478,0.351095)
 (0.714478,0.350234)
 (0.714478,0.349378)
 (0.714478,0.348526)
 (0.714478,0.347678)
 (0.714478,0.346834)
 (0.714478,0.345994)
 (0.714478,0.345158)
 (0.714478,0.344327)
 (0.714478,0.343499)
 (0.714478,0.342675)
 (0.714478,0.341855)
 (0.714478,0.341039)
 (0.714478,0.340227)
 (0.714478,0.339419)
 (0.714478,0.338615)
 (0.714478,0.337814)
 (0.714478,0.337018)
 (0.714478,0.336225)
 (0.714478,0.335435)
 (0.714478,0.33465)
 (0.714478,0.333868)
 (0.714478,0.33309)
 (0.714478,0.332315)
 (0.714478,0.331544)
 (0.714478,0.330777)
 (0.714478,0.330013)
 (0.714478,0.329252)
 (0.714478,0.328495)
 (0.714478,0.327742)
 (0.714478,0.326992)
 (0.714478,0.326245)
 (0.714478,0.325502)
 (0.714478,0.324763)
 (0.714478,0.324026)
 (0.714478,0.323293)
 (0.714478,0.322563)
 (0.714478,0.321837)
 (0.714478,0.321114)
 (0.714478,0.320394)
 (0.714478,0.319677)
 (0.714478,0.318963)
 (0.714478,0.318253)
 (0.714478,0.317546)
 (0.714478,0.316841)
 (0.714478,0.316141)
 (0.714478,0.315443)
 (0.714478,0.314748)
 (0.714478,0.314056)
 (0.714478,0.313367)
 (0.714478,0.312682)
 (0.714478,0.311999)
 (0.714478,0.311319)
 (0.714478,0.310642)
 (0.714478,0.309969)
 (0.714478,0.309298)
 (0.714478,0.30863)
 (0.714478,0.307964)
 (0.714478,0.307302)
 (0.714478,0.306643)
 (0.714478,0.305986)
 (0.714478,0.305332)
 (0.714478,0.304681)
 (0.714478,0.304033)
 (0.714478,0.303387)
 (0.714478,0.302745)
 (0.714478,0.302105)
 (0.714478,0.301467)
 (0.714478,0.300833)
 (0.714478,0.300201)
 (0.714478,0.299571)
 (0.714478,0.298945)
 (0.714478,0.29832)
 (0.714478,0.297699)
 (0.714478,0.29708)
 (0.714478,0.296464)
 (0.714478,0.29585)
 (0.714478,0.295239)
 (0.714478,0.29463)
 (0.714478,0.294024)
 (0.714478,0.29342)
 (0.714478,0.292819)
 (0.714478,0.29222)
 (0.714478,0.291623)
 (0.714478,0.29103)
 (0.714478,0.290438)
 (0.714478,0.289849)
 (0.714478,0.289262)
 (0.714478,0.288678)
 (0.714478,0.288096)
 (0.714478,0.287516)
 (0.714478,0.286939)
 (0.714478,0.286364)
 (0.714478,0.285791)
 (0.714478,0.285221)
 (0.714478,0.284652)
 (0.714478,0.284086)
 (0.714478,0.283523)
 (0.714478,0.282961)
 (0.714478,0.282402)
 (0.714478,0.281845)
 (0.714478,0.28129)
 (0.714478,0.280738)
 (0.714478,0.280187)
 (0.714478,0.279639)
 (0.714478,0.279093)
 (0.714478,0.278549)
 (0.714478,0.278007)
 (0.714478,0.277467)
 (0.714478,0.276929)
 (0.714478,0.276394)
 (0.714478,0.27586)
 (0.714478,0.275329)
 (0.714478,0.274799)
 (0.714478,0.274272)
 (0.714478,0.273746)
 (0.714478,0.273223)
 (0.714478,0.272701)
 (0.714478,0.272182)
 (0.714478,0.271664)
 (0.714478,0.271149)
 (0.714478,0.270635)
 (0.714478,0.270124)
 (0.714478,0.269614)
 (0.714478,0.269106)
 (0.714478,0.268601)
 (0.714478,0.268097)
 (0.714478,0.267595)
 (0.714478,0.267094)
 (0.714478,0.266596)
 (0.714478,0.2661)
 (0.714478,0.265605)
 (0.714478,0.265112)
 (0.714478,0.264621)
 (0.714478,0.264132)
 (0.714478,0.263645)
 (0.714478,0.263159)
 (0.714478,0.262676)
 (0.714478,0.262194)
 (0.714478,0.261713)
 (0.714478,0.261235)
 (0.714478,0.260758)
 (0.714478,0.260283)
 (0.714478,0.25981)
 (0.714478,0.259338)
 (0.714478,0.258869)
 (0.714478,0.258401)
 (0.714478,0.257934)
 (0.714478,0.257469)
 (0.714478,0.257006)
 (0.714478,0.256545)
 (0.714478,0.256085)
 (0.714478,0.255627)
 (0.714478,0.255171)
 (0.714478,0.254716)
 (0.714478,0.254262)
 (0.714478,0.253811)
 (0.714478,0.253361)
 (0.714478,0.252912)
 (0.714478,0.252466)
 (0.714478,0.25202)
 (0.714478,0.251577)
 (0.714478,0.251134)
 (0.714478,0.250694)
 (0.714478,0.250255)
 (0.714478,0.249817)
 (0.714478,0.249381)
 (0.714478,0.248947)
 (0.714478,0.248514)
 (0.714478,0.248082)
 (0.714478,0.247653)
 (0.714478,0.247224)
 (0.714478,0.246797)
 (0.714478,0.246372)
 (0.714478,0.245948)
 (0.714478,0.245525)
 (0.714478,0.245104)
 (0.714478,0.244684)
 (0.714478,0.244266)
 (0.714478,0.243849)
 (0.714478,0.243434)
 (0.714478,0.24302)
 (0.714478,0.242607)
 (0.714478,0.242196)
 (0.714478,0.241786)
 (0.714478,0.241378)
 (0.714478,0.240971)
 (0.714478,0.240565)
 (0.714478,0.240161)
 (0.714478,0.239758)
 (0.714478,0.239356)
 (0.714478,0.238956)
 (0.714478,0.238557)
 (0.714478,0.238159)
 (0.714478,0)
} |- (axis cs:0,0) -- cycle;

\addplot [
color=blue,
solid,
line width=1.1pt
]coordinates{
 (0,0.71) (0.49,0.71) (0.5,0.7) (0.5,0.68) (0.51,0.67) (0.51,0.65) (0.52,0.65) 
 (0.52,0.64) (0.53,0.63) (0.53,0.62) (0.53,0.61) (0.54,0.61) (0.54,0.60) (0.55,0.59)
 (0.56,0.58) (0.57,0.57) (0.58,0.56) (0.59,0.55) (0.60,0.54) (0.61,0.54) (0.61,0.53)
 (0.62,0.53) (0.63,0.53) (0.64,0.52) (0.65,0.52) (0.65,0.51) (0.67,0.51) (0.68,0.50)
 (0.70,0.50) (0.71,0.49) (0.71,0)}
|- (axis cs:0,0) -- cycle;

\addplot [
color=red,
solid,
line width=1.5pt
]
coordinates{
 (0.7519,0) (0.75,0.5037) (0.5037,0.7519) (0,0.7519)
};


\end{axis}

\end{tikzpicture}

%% file: awgn_opt_roc.tex
%

\begin{tikzpicture}[scale=0.47]

\begin{axis}[normalsize,
scale only axis,
width=4.5in,
height=4.5in,
xmin=-6, xmax=3,
ymin=-6, ymax=3,
xtick={-6,-5,-4,-3,-2,-1,0,1,2,3},
ytick={-6,-5,-4,-3,-2,-1,0,1,2,3},
xlabel={\Large$\text{SNR}_1$ (dB)},
ylabel={\Large$\text{SNR}_2$ (dB)},
xmajorgrids,
ymajorgrids]

\addplot [
color=red,
solid,
line width=1.5pt
]
coordinates{
  (-4.39,3)
 (-4.39,-0.29)
 (-4.38,-0.29)
 (-4.37,-0.3)
 (-4.36,-0.3)
 (-4.35,-0.31)
 (-4.34,-0.32)
 (-4.33,-0.32)
 (-4.32,-0.33)
 (-4.31,-0.33)
 (-4.3,-0.34)
 (-4.29,-0.34)
 (-4.28,-0.35)
 (-4.27,-0.35)
 (-4.26,-0.36)
 (-4.25,-0.37)
 (-4.24,-0.37)
 (-4.23,-0.38)
 (-4.22,-0.38)
 (-4.21,-0.39)
 (-4.2,-0.39)
 (-4.19,-0.4)
 (-4.18,-0.41)
 (-4.17,-0.41)
 (-4.16,-0.42)
 (-4.15,-0.42)
 (-4.14,-0.43)
 (-4.13,-0.44)
 (-4.12,-0.44)
 (-4.11,-0.45)
 (-4.1,-0.45)
 (-4.09,-0.46)
 (-4.08,-0.46)
 (-4.07,-0.47)
 (-4.06,-0.48)
 (-4.05,-0.48)
 (-4.04,-0.49)
 (-4.03,-0.49)
 (-4.02,-0.5)
 (-4.01,-0.51)
 (-4,-0.51)
 (-3.99,-0.52)
 (-3.98,-0.53)
 (-3.97,-0.53)
 (-3.96,-0.54)
 (-3.95,-0.54)
 (-3.94,-0.55)
 (-3.93,-0.56)
 (-3.92,-0.56)
 (-3.91,-0.57)
 (-3.9,-0.57)
 (-3.89,-0.58)
 (-3.88,-0.59)
 (-3.87,-0.59)
 (-3.86,-0.6)
 (-3.85,-0.61)
 (-3.84,-0.61)
 (-3.83,-0.62)
 (-3.82,-0.62)
 (-3.81,-0.63)
 (-3.8,-0.64)
 (-3.79,-0.64)
 (-3.78,-0.65)
 (-3.77,-0.66)
 (-3.76,-0.66)
 (-3.75,-0.67)
 (-3.74,-0.68)
 (-3.73,-0.68)
 (-3.72,-0.69)
 (-3.71,-0.69)
 (-3.7,-0.7)
 (-3.69,-0.71)
 (-3.68,-0.71)
 (-3.67,-0.72)
 (-3.66,-0.73)
 (-3.65,-0.73)
 (-3.64,-0.74)
 (-3.63,-0.75)
 (-3.62,-0.75)
 (-3.61,-0.76)
 (-3.6,-0.77)
 (-3.59,-0.77)
 (-3.58,-0.78)
 (-3.57,-0.79)
 (-3.56,-0.79)
 (-3.55,-0.8)
 (-3.54,-0.81)
 (-3.53,-0.81)
 (-3.52,-0.82)
 (-3.51,-0.83)
 (-3.5,-0.83)
 (-3.49,-0.84)
 (-3.48,-0.85)
 (-3.47,-0.86)
 (-3.46,-0.86)
 (-3.45,-0.87)
 (-3.44,-0.88)
 (-3.43,-0.88)
 (-3.42,-0.89)
 (-3.41,-0.9)
 (-3.4,-0.9)
 (-3.39,-0.91)
 (-3.38,-0.92)
 (-3.37,-0.92)
 (-3.36,-0.93)
 (-3.35,-0.94)
 (-3.34,-0.95)
 (-3.33,-0.95)
 (-3.32,-0.96)
 (-3.31,-0.97)
 (-3.3,-0.97)
 (-3.29,-0.98)
 (-3.28,-0.99)
 (-3.27,-1)
 (-3.26,-1)
 (-3.25,-1.01)
 (-3.24,-1.02)
 (-3.23,-1.03)
 (-3.22,-1.03)
 (-3.21,-1.04)
 (-3.2,-1.05)
 (-3.19,-1.05)
 (-3.18,-1.06)
 (-3.17,-1.07)
 (-3.16,-1.08)
 (-3.15,-1.08)
 (-3.14,-1.09)
 (-3.13,-1.1)
 (-3.12,-1.11)
 (-3.11,-1.11)
 (-3.1,-1.12)
 (-3.09,-1.13)
 (-3.08,-1.14)
 (-3.07,-1.14)
 (-3.06,-1.15)
 (-3.05,-1.16)
 (-3.04,-1.17)
 (-3.03,-1.18)
 (-3.02,-1.18)
 (-3.01,-1.19)
 (-3,-1.2)
 (-2.99,-1.21)
 (-2.98,-1.21)
 (-2.97,-1.22)
 (-2.96,-1.23)
 (-2.95,-1.24)
 (-2.94,-1.25)
 (-2.93,-1.25)
 (-2.92,-1.26)
 (-2.91,-1.27)
 (-2.9,-1.28)
 (-2.89,-1.28)
 (-2.88,-1.29)
 (-2.87,-1.3)
 (-2.86,-1.31)
 (-2.85,-1.32)
 (-2.84,-1.33)
 (-2.83,-1.33)
 (-2.82,-1.34)
 (-2.81,-1.35)
 (-2.8,-1.36)
 (-2.79,-1.37)
 (-2.78,-1.37)
 (-2.77,-1.38)
 (-2.76,-1.39)
 (-2.75,-1.4)
 (-2.74,-1.41)
 (-2.73,-1.42)
 (-2.72,-1.42)
 (-2.71,-1.43)
 (-2.7,-1.44)
 (-2.69,-1.45)
 (-2.68,-1.46)
 (-2.67,-1.47)
 (-2.66,-1.47)
 (-2.65,-1.48)
 (-2.64,-1.49)
 (-2.63,-1.5)
 (-2.62,-1.51)
 (-2.61,-1.52)
 (-2.6,-1.52)
 (-2.59,-1.53)
 (-2.58,-1.54)
 (-2.57,-1.55)
 (-2.56,-1.56)
 (-2.55,-1.57)
 (-2.54,-1.58)
 (-2.53,-1.59)
 (-2.52,-1.59)
 (-2.51,-1.6)
 (-2.5,-1.61)
 (-2.49,-1.62)
 (-2.48,-1.63)
 (-2.47,-1.64)
 (-2.46,-1.65)
 (-2.45,-1.66)
 (-2.44,-1.67)
 (-2.43,-1.67)
 (-2.42,-1.68)
 (-2.41,-1.69)
 (-2.4,-1.7)
 (-2.39,-1.71)
 (-2.38,-1.72)
 (-2.37,-1.73)
 (-2.36,-1.74)
 (-2.35,-1.75)
 (-2.34,-1.76)
 (-2.33,-1.77)
 (-2.32,-1.77)
 (-2.31,-1.78)
 (-2.3,-1.79)
 (-2.29,-1.8)
 (-2.28,-1.81)
 (-2.27,-1.82)
 (-2.26,-1.83)
 (-2.25,-1.84)
 (-2.24,-1.85)
 (-2.23,-1.86)
 (-2.22,-1.87)
 (-2.21,-1.88)
 (-2.2,-1.89)
 (-2.19,-1.9)
 (-2.18,-1.91)
 (-2.17,-1.92)
 (-2.16,-1.93)
 (-2.15,-1.94)
 (-2.14,-1.95)
 (-2.13,-1.96)
 (-2.12,-1.97)
 (-2.11,-1.97)
 (-2.1,-1.98)
 (-2.09,-1.99)
 (-2.08,-2)
 (-2.07,-2.01)
 (-2.06,-2.02)
 (-2.05,-2.03)
 (-2.04,-2.04)
 (-2.03,-2.05)
 (-2.02,-2.06)
 (-2.01,-2.07)
 (-2,-2.08)
 (-1.99,-2.09)
 (-1.98,-2.1)
 (-1.97,-2.11)
 (-1.97,-2.12)
 (-1.96,-2.13)
 (-1.95,-2.14)
 (-1.94,-2.15)
 (-1.93,-2.16)
 (-1.92,-2.17)
 (-1.91,-2.18)
 (-1.9,-2.19)
 (-1.89,-2.2)
 (-1.88,-2.21)
 (-1.87,-2.22)
 (-1.86,-2.23)
 (-1.85,-2.24)
 (-1.84,-2.25)
 (-1.83,-2.26)
 (-1.82,-2.27)
 (-1.81,-2.28)
 (-1.8,-2.29)
 (-1.79,-2.3)
 (-1.78,-2.31)
 (-1.77,-2.32)
 (-1.77,-2.33)
 (-1.76,-2.34)
 (-1.75,-2.35)
 (-1.74,-2.36)
 (-1.73,-2.37)
 (-1.72,-2.38)
 (-1.71,-2.39)
 (-1.7,-2.4)
 (-1.69,-2.41)
 (-1.68,-2.42)
 (-1.67,-2.43)
 (-1.67,-2.44)
 (-1.66,-2.45)
 (-1.65,-2.46)
 (-1.64,-2.47)
 (-1.63,-2.48)
 (-1.62,-2.49)
 (-1.61,-2.5)
 (-1.6,-2.51)
 (-1.59,-2.52)
 (-1.59,-2.53)
 (-1.58,-2.54)
 (-1.57,-2.55)
 (-1.56,-2.56)
 (-1.55,-2.57)
 (-1.54,-2.58)
 (-1.53,-2.59)
 (-1.52,-2.6)
 (-1.52,-2.61)
 (-1.51,-2.62)
 (-1.5,-2.63)
 (-1.49,-2.64)
 (-1.48,-2.65)
 (-1.47,-2.66)
 (-1.47,-2.67)
 (-1.46,-2.68)
 (-1.45,-2.69)
 (-1.44,-2.7)
 (-1.43,-2.71)
 (-1.42,-2.72)
 (-1.42,-2.73)
 (-1.41,-2.74)
 (-1.4,-2.75)
 (-1.39,-2.76)
 (-1.38,-2.77)
 (-1.37,-2.78)
 (-1.37,-2.79)
 (-1.36,-2.8)
 (-1.35,-2.81)
 (-1.34,-2.82)
 (-1.33,-2.83)
 (-1.33,-2.84)
 (-1.32,-2.85)
 (-1.31,-2.86)
 (-1.3,-2.87)
 (-1.29,-2.88)
 (-1.28,-2.89)
 (-1.28,-2.9)
 (-1.27,-2.91)
 (-1.26,-2.92)
 (-1.25,-2.93)
 (-1.25,-2.94)
 (-1.24,-2.95)
 (-1.23,-2.96)
 (-1.22,-2.97)
 (-1.21,-2.98)
 (-1.21,-2.99)
 (-1.2,-3)
 (-1.19,-3.01)
 (-1.18,-3.02)
 (-1.18,-3.03)
 (-1.17,-3.04)
 (-1.16,-3.05)
 (-1.15,-3.06)
 (-1.14,-3.07)
 (-1.14,-3.08)
 (-1.13,-3.09)
 (-1.12,-3.1)
 (-1.11,-3.11)
 (-1.11,-3.12)
 (-1.1,-3.13)
 (-1.09,-3.14)
 (-1.08,-3.15)
 (-1.08,-3.16)
 (-1.07,-3.17)
 (-1.06,-3.18)
 (-1.05,-3.19)
 (-1.05,-3.2)
 (-1.04,-3.21)
 (-1.03,-3.22)
 (-1.03,-3.23)
 (-1.02,-3.24)
 (-1.01,-3.25)
 (-1,-3.26)
 (-1,-3.27)
 (-0.99,-3.28)
 (-0.98,-3.29)
 (-0.97,-3.3)
 (-0.97,-3.31)
 (-0.96,-3.32)
 (-0.95,-3.33)
 (-0.95,-3.34)
 (-0.94,-3.35)
 (-0.93,-3.36)
 (-0.92,-3.37)
 (-0.92,-3.38)
 (-0.91,-3.39)
 (-0.9,-3.4)
 (-0.9,-3.41)
 (-0.89,-3.42)
 (-0.88,-3.43)
 (-0.88,-3.44)
 (-0.87,-3.45)
 (-0.86,-3.46)
 (-0.86,-3.47)
 (-0.85,-3.48)
 (-0.84,-3.49)
 (-0.83,-3.5)
 (-0.83,-3.51)
 (-0.82,-3.52)
 (-0.81,-3.53)
 (-0.81,-3.54)
 (-0.8,-3.55)
 (-0.79,-3.56)
 (-0.79,-3.57)
 (-0.78,-3.58)
 (-0.77,-3.59)
 (-0.77,-3.6)
 (-0.76,-3.61)
 (-0.75,-3.62)
 (-0.75,-3.63)
 (-0.74,-3.64)
 (-0.73,-3.65)
 (-0.73,-3.66)
 (-0.72,-3.67)
 (-0.71,-3.68)
 (-0.71,-3.69)
 (-0.7,-3.7)
 (-0.69,-3.71)
 (-0.69,-3.72)
 (-0.68,-3.73)
 (-0.68,-3.74)
 (-0.67,-3.75)
 (-0.66,-3.76)
 (-0.66,-3.77)
 (-0.65,-3.78)
 (-0.64,-3.79)
 (-0.64,-3.8)
 (-0.63,-3.81)
 (-0.62,-3.82)
 (-0.62,-3.83)
 (-0.61,-3.84)
 (-0.61,-3.85)
 (-0.6,-3.86)
 (-0.59,-3.87)
 (-0.59,-3.88)
 (-0.58,-3.89)
 (-0.57,-3.9)
 (-0.57,-3.91)
 (-0.56,-3.92)
 (-0.56,-3.93)
 (-0.55,-3.94)
 (-0.54,-3.95)
 (-0.54,-3.96)
 (-0.53,-3.97)
 (-0.53,-3.98)
 (-0.52,-3.99)
 (-0.51,-4)
 (-0.51,-4.01)
 (-0.5,-4.02)
 (-0.49,-4.03)
 (-0.49,-4.04)
 (-0.48,-4.05)
 (-0.48,-4.06)
 (-0.47,-4.07)
 (-0.46,-4.08)
 (-0.46,-4.09)
 (-0.45,-4.1)
 (-0.45,-4.11)
 (-0.44,-4.12)
 (-0.44,-4.13)
 (-0.43,-4.14)
 (-0.42,-4.15)
 (-0.42,-4.16)
 (-0.41,-4.17)
 (-0.41,-4.18)
 (-0.4,-4.19)
 (-0.39,-4.2)
 (-0.39,-4.21)
 (-0.38,-4.22)
 (-0.38,-4.23)
 (-0.37,-4.24)
 (-0.37,-4.25)
 (-0.36,-4.26)
 (-0.35,-4.27)
 (-0.35,-4.28)
 (-0.34,-4.29)
 (-0.34,-4.3)
 (-0.33,-4.31)
 (-0.33,-4.32)
 (-0.32,-4.33)
 (-0.32,-4.34)
 (-0.31,-4.35)
 (-0.3,-4.36)
 (-0.3,-4.37)
 (-0.29,-4.38)
 (-0.29,-4.39)
 (-0.28,-4.39)
 (3,-4.39)

};

\addplot [
color=red,
solid
]
coordinates{
 (-4.39,2.6)
 (-4.39,2.59)
 (-4.39,2.58)
 (-4.39,2.57)
 (-4.39,2.56)
 (-4.39,2.55)
 (-4.39,2.54)
 (-4.39,2.53)
 (-4.39,2.52)
 (-4.39,2.51)
 (-4.39,2.5)
 (-4.39,2.49)
 (-4.39,2.48)
 (-4.39,2.47)
 (-4.39,2.46)
 (-4.39,2.45)
 (-4.39,2.44)
 (-4.39,2.43)
 (-4.39,2.42)
 (-4.39,2.41)
 (-4.39,2.4)
 (-4.39,2.39)
 (-4.39,2.38)
 (-4.39,2.37)
 (-4.39,2.36)
 (-4.39,2.35)
 (-4.39,2.34)
 (-4.39,2.33)
 (-4.39,2.32)
 (-4.39,2.31)
 (-4.39,2.3)
 (-4.39,2.29)
 (-4.39,2.28)
 (-4.39,2.27)
 (-4.39,2.26)
 (-4.39,2.25)
 (-4.39,2.24)
 (-4.39,2.23)
 (-4.39,2.22)
 (-4.39,2.21)
 (-4.39,2.2)
 (-4.39,2.19)
 (-4.39,2.18)
 (-4.39,2.17)
 (-4.39,2.16)
 (-4.39,2.15)
 (-4.39,2.14)
 (-4.39,2.13)
 (-4.39,2.12)
 (-4.39,2.11)
 (-4.39,2.1)
 (-4.39,2.09)
 (-4.39,2.08)
 (-4.39,2.07)
 (-4.39,2.06)
 (-4.39,2.05)
 (-4.39,2.04)
 (-4.39,2.03)
 (-4.39,2.02)
 (-4.39,2.01)
 (-4.39,2)
 (-4.39,1.99)
 (-4.39,1.98)
 (-4.39,1.97)
 (-4.39,1.96)
 (-4.39,1.95)
 (-4.39,1.94)
 (-4.39,1.93)
 (-4.39,1.92)
 (-4.39,1.91)
 (-4.39,1.9)
 (-4.39,1.89)
 (-4.39,1.88)
 (-4.39,1.87)
 (-4.39,1.86)
 (-4.39,1.85)
 (-4.39,1.84)
 (-4.39,1.83)
 (-4.39,1.82)
 (-4.39,1.81)
 (-4.39,1.8)
 (-4.39,1.79)
 (-4.39,1.78)
 (-4.39,1.77)
 (-4.39,1.76)
 (-4.39,1.75)
 (-4.39,1.74)
 (-4.39,1.73)
 (-4.39,1.72)
 (-4.39,1.71)
 (-4.39,1.7)
 (-4.39,1.69)
 (-4.39,1.68)
 (-4.39,1.67)
 (-4.39,1.66)
 (-4.39,1.65)
 (-4.39,1.64)
 (-4.39,1.63)
 (-4.39,1.62)
 (-4.39,1.61)
 (-4.39,1.6)
 (-4.39,1.59)
 (-4.39,1.58)
 (-4.39,1.57)
 (-4.39,1.56)
 (-4.39,1.55)
 (-4.39,1.54)
 (-4.39,1.53)
 (-4.39,1.52)
 (-4.39,1.51)
 (-4.39,1.5)
 (-4.39,1.49)
 (-4.39,1.48)
 (-4.39,1.47)
 (-4.39,1.46)
 (-4.39,1.45)
 (-4.39,1.44)
 (-4.39,1.43)
 (-4.39,1.42)
 (-4.39,1.41)
 (-4.39,1.4)
 (-4.39,1.39)
 (-4.39,1.38)
 (-4.39,1.37)
 (-4.39,1.36)
 (-4.39,1.35)
 (-4.39,1.34)
 (-4.39,1.33)
 (-4.39,1.32)
 (-4.39,1.31)
 (-4.39,1.3)
 (-4.39,1.29)
 (-4.39,1.28)
 (-4.39,1.27)
 (-4.39,1.26)
 (-4.39,1.25)
 (-4.39,1.24)
 (-4.39,1.23)
 (-4.39,1.22)
 (-4.39,1.21)
 (-4.39,1.2)
 (-4.39,1.19)
 (-4.39,1.18)
 (-4.39,1.17)
 (-4.39,1.16)
 (-4.39,1.15)
 (-4.39,1.14)
 (-4.39,1.13)
 (-4.39,1.12)
 (-4.39,1.11)
 (-4.39,1.1)
 (-4.39,1.09)
 (-4.39,1.08)
 (-4.39,1.07)
 (-4.39,1.06)
 (-4.39,1.05)
 (-4.39,1.04)
 (-4.39,1.03)
 (-4.39,1.02)
 (-4.39,1.01)
 (-4.39,1)
 (-4.39,0.99)
 (-4.39,0.98)
 (-4.39,0.97)
 (-4.39,0.96)
 (-4.39,0.95)
 (-4.39,0.94)
 (-4.39,0.93)
 (-4.39,0.92)
 (-4.39,0.91)
 (-4.39,0.9)
 (-4.39,0.89)
 (-4.39,0.88)
 (-4.39,0.87)
 (-4.39,0.86)
 (-4.39,0.85)
 (-4.39,0.84)
 (-4.39,0.83)
 (-4.39,0.82)
 (-4.39,0.81)
 (-4.39,0.8)
 (-4.39,0.79)
 (-4.39,0.78)
 (-4.39,0.77)
 (-4.39,0.76)
 (-4.39,0.75)
 (-4.39,0.74)
 (-4.39,0.73)
 (-4.39,0.72)
 (-4.39,0.71)
 (-4.39,0.7)
 (-4.39,0.69)
 (-4.39,0.68)
 (-4.39,0.67)
 (-4.39,0.66)
 (-4.39,0.65)
 (-4.39,0.64)
 (-4.39,0.63)
 (-4.39,0.62)
 (-4.39,0.61)
 (-4.39,0.6)
 (-4.39,0.59)
 (-4.39,0.58)
 (-4.39,0.57)
 (-4.39,0.56)
 (-4.39,0.55)
 (-4.39,0.54)
 (-4.39,0.53)
 (-4.39,0.52)
 (-4.39,0.51)
 (-4.39,0.5)
 (-4.39,0.49)
 (-4.39,0.48)
 (-4.39,0.47)
 (-4.39,0.46)
 (-4.39,0.45)
 (-4.39,0.44)
 (-4.39,0.43)
 (-4.39,0.42)
 (-4.39,0.41)
 (-4.39,0.4)
 (-4.39,0.39)
 (-4.39,0.38)
 (-4.39,0.37)
 (-4.39,0.36)
 (-4.39,0.35)
 (-4.39,0.34)
 (-4.39,0.33)
 (-4.39,0.32)
 (-4.39,0.31)
 (-4.39,0.3)
 (-4.39,0.29)
 (-4.39,0.28)
 (-4.39,0.27)
 (-4.39,0.26)
 (-4.39,0.25)
 (-4.39,0.24)
 (-4.39,0.23)
 (-4.39,0.22)
 (-4.39,0.21)
 (-4.39,0.2)
 (-4.39,0.19)
 (-4.39,0.18)
 (-4.39,0.17)
 (-4.39,0.16)
 (-4.39,0.15)
 (-4.39,0.14)
 (-4.39,0.13)
 (-4.39,0.12)
 (-4.39,0.11)
 (-4.39,0.1)
 (-4.39,0.09)
 (-4.39,0.08)
 (-4.39,0.07)
 (-4.39,0.06)
 (-4.39,0.05)
 (-4.39,0.04)
 (-4.39,0.03)
 (-4.39,0.02)
 (-4.39,0.01)
 (-4.39,0)
 (-4.39,-0.01)
 (-4.39,-0.02)
 (-4.39,-0.03)
 (-4.39,-0.04)
 (-4.39,-0.05)
 (-4.39,-0.06)
 (-4.39,-0.07)
 (-4.39,-0.08)
 (-4.39,-0.09)
 (-4.39,-0.1)
 (-4.39,-0.11)
 (-4.39,-0.12)
 (-4.39,-0.13)
 (-4.39,-0.14)
 (-4.39,-0.15)
 (-4.39,-0.16)
 (-4.39,-0.17)
 (-4.39,-0.18)
 (-4.39,-0.19)
 (-4.39,-0.2)
 (-4.39,-0.21)
 (-4.39,-0.22)
 (-4.39,-0.23)
 (-4.39,-0.24)
 (-4.39,-0.25)
 (-4.39,-0.26)
 (-4.39,-0.27)
 (-4.39,-0.28)
 (-4.39,-0.29)

};

\addplot
[black,fill=gray,fill opacity=0.4]
coordinates{
 (-3.56055,3)
 (-3.56055,-0.0839844)
 (-3.35116,-0.183984)
 (-3.06458,-0.283984)
 (-2.84163,-0.383984)
 (-2.65996,-0.483984)
 (-2.50829,-0.583984)
 (-2.38352,-0.683984)
 (-2.2753,-0.783984)
 (-2.18175,-0.883984)
 (-2.09613,-0.983984)
 (-2.02237,-1.08398)
 (-1.96408,-1.18398)
 (-1.89759,-1.28398)
 (-1.85161,-1.38398)
 (-1.80034,-1.48398)
 (-1.74226,-1.58398)
 (-1.58398,-1.74226)
 (-1.48398,-1.80034)
 (-1.38398,-1.85161)
 (-1.28398,-1.89759)
 (-1.18398,-1.96408)
 (-1.08398,-2.02237)
 (-0.983984,-2.09613)
 (-0.883984,-2.18175)
 (-0.783984,-2.2753)
 (-0.683984,-2.38352)
 (-0.583984,-2.50829)
 (-0.483984,-2.65996)
 (-0.383984,-2.84163)
 (-0.283984,-3.06458)
 (-0.183984,-3.35116)
 (-0.0839844,-3.56055)
 (3,-3.56055)

}|- (axis cs:3,3) -- cycle;

\addplot
[blue,solid,line width=1.1pt]
coordinates{
  (-3.35,3)
 (-3.35,0.6)
 (-3.3,0.55)
 (-3.3,0.5)
 (-3.3,0.45)
 (-3.3,0.4)
 (-3.3,0.35)
 (-3.3,0.3)
 (-3.25,0.25)
 (-3.25,0.2)
 (-3.25,0.15)
 (-3.2,0.1)
 (-3.15,0.05)
 (-3.1,0.05)
 (-3.05,0)
 (-3,0)
 (-2.95,0)
 (-2.9,-0.05)
 (-2.85,-0.05)
 (-2.8,-0.05)
 (-2.75,-0.1)
 (-2.7,-0.1)
 (-2.65,-0.1)
 (-2.6,-0.1)
 (-2.55,-0.15)
 (-2.5,-0.15)
 (-2.45,-0.15)
 (-2.4,-0.2)
 (-2.35,-0.2)
 (-2.3,-0.2)
 (-2.25,-0.25)
 (-2.2,-0.25)
 (-2.15,-0.25)
 (-2.1,-0.3)
 (-2.05,-0.3)
 (-2,-0.35)
 (-1.95,-0.35)
 (-1.9,-0.35)
 (-1.85,-0.4)
 (-1.8,-0.4)
 (-1.75,-0.45)
 (-1.7,-0.45)
 (-1.65,-0.5)
 (-1.6,-0.5)
 (-1.55,-0.55)
 (-1.5,-0.55)
 (-1.45,-0.6)
 (-1.4,-0.65)
 (-1.35,-0.65)
 (-1.3,-0.7)
 (-1.25,-0.75)
 (-1.2,-0.8)
 (-1.15,-0.8)
 (-1.1,-0.85)
 (-1.05,-0.9)
 (-1,-0.95)
 (-0.95,-1)
 (-0.9,-1.05)
 (-0.85,-1.1)
 (-0.8,-1.15)
 (-0.8,-1.2)
 (-0.75,-1.25)
 (-0.7,-1.3)
 (-0.65,-1.35)
 (-0.65,-1.4)
 (-0.6,-1.45)
 (-0.55,-1.5)
 (-0.55,-1.55)
 (-0.5,-1.6)
 (-0.5,-1.65)
 (-0.45,-1.7)
 (-0.45,-1.75)
 (-0.4,-1.8)
 (-0.4,-1.85)
 (-0.35,-1.9)
 (-0.35,-1.95)
 (-0.35,-2)
 (-0.3,-2.05)
 (-0.3,-2.1)
 (-0.25,-2.15)
 (-0.25,-2.2)
 (-0.25,-2.25)
 (-0.2,-2.3)
 (-0.2,-2.35)
 (-0.2,-2.4)
 (-0.15,-2.45)
 (-0.15,-2.5)
 (-0.15,-2.55)
 (-0.1,-2.6)
 (-0.1,-2.65)
 (-0.1,-2.7)
 (-0.1,-2.75)
 (-0.05,-2.8)
 (-0.05,-2.85)
 (-0.05,-2.9)
 (0,-2.95)
 (0,-3)
 (0,-3.05)
 (0.05,-3.1)
 (0.05,-3.15)
 (0.1,-3.2)
 (0.15,-3.25)
 (0.2,-3.25)
 (0.25,-3.25)
 (0.3,-3.3)
 (0.35,-3.3)
 (0.4,-3.3)
 (0.45,-3.3)
 (0.5,-3.3)
 (0.55,-3.3)
 (0.6,-3.3)
 (0.65,-3.35)
 (0.7,-3.35)
 (0.75,-3.35)
 (0.8,-3.35)
 (0.85,-3.35)
 (0.9,-3.35)
 (0.95,-3.35)
 (1,-3.35)
 (1.05,-3.35)
 (1.1,-3.35)
 (1.15,-3.35)
 (1.2,-3.35)
 (1.25,-3.35)
 (1.3,-3.35)
 (1.35,-3.35)
 (1.4,-3.35)
 (1.45,-3.35)
 (1.5,-3.35)
 (1.55,-3.35)
 (1.6,-3.35)
  (3,-3.35)
}|- (axis cs:3,3) -- cycle;
\end{axis}

\end{tikzpicture}

%% file: appendix.tex
\section{Proof of Theorem \ref{thm-sum-rate}}
\label{sec:proof-theorem-refthm}
We will use the following Lemma to show that the density evolution
equations converge to zero at the extremal symmetric point.
\begin{lem}
  \label{sum-rate}
  \begin{equation*}
    \rho^N(x)> \frac{\mu+\rho(x)}{\mu+G_N(p)+1}, \text{ for } 0\leq x<1-\frac{1}{N}.
  \end{equation*}
\end{lem} 
\begin{IEEEproof}
  For $0\leq x<1-\frac{1}{N}$, we have
  \begin{align}
    \rho^N(x) &= \frac{\mu+\sum_{i=1}^{N} \frac{\sum_{k=0}^{i-1}\binom{2i-1}{k}p^k}{i(1+p)^{2i-1}}x^i+x^N}{\mu+G_N(p)+1}\notag\\
    &= \frac{\mu+\rho(x)+x^N}{\mu+G_N(p)+1}-
    \frac{\sum_{i=N+1}^{\infty} \frac{\sum_{k=0}^{i-1}\binom{2i-1}{k}p^k}{i(1+p)^{2i-1}}x^i}{\mu+G_N(p)+1}\notag\\
    &> \frac{\mu+\rho(x)}{\mu+G_N(p)+1}.\label{eq:ineq1}\\
    \intertext{ \eqref{eq:ineq1} follows from the fact that}
    \sum_{i=N+1}^{\infty} \frac{\sum_{k=0}^{i-1}\binom{2i-1}{k}p^k}{i(1+p)^{2i-1}}x^i &< \sum_{i=N+1}^{\infty}\frac{x^i}{i}
    < \frac{1}{N+1}\sum_{i=N+1}^{\infty}x^i
    = \frac{1}{N+1}\cdot\frac{x^{N+1}}{1-x}
    < x^N.\notag
  \end{align}
  The last step follows from explicit calculations, taking into
  account that $0\leq x<1-\frac{1}{N}$.      
\end{IEEEproof}
From \eqref{eq:one-de}, the convergence criteria for the density
evolution equation is given by
\begin{equation*}
  \begin{split}
    x >  \left[(1-p)+p\bar{\lambda}^N(\epsilon,x)\right]\bar{\lambda}^N(\epsilon,x),
  \end{split}
\end{equation*}
where $\bar{\lambda}^N(\epsilon,x) =
\lambda\left(1-(1-\epsilon)\rho^N(1-x)\right)$. We have,
\begin{align}
    \bar{\lambda}^N(\epsilon,x) &= e^{-\mathsf{m}(1-\epsilon)\cdot\rho^N(1-x)}\notag\\
    &\leq e^{-\mathsf{m}(1-\epsilon)\frac{\mu+\rho(1-x)}{\mu+G_N(p)+1}},\textnormal{ if } x\geq\frac{1}{N}\label{eq:ineq3}\\
    &< e^{-\mu}\cdot \frac{\sqrt{(1-p)^2+4px}-(1-p)}{2p}\notag\\
    &< \frac{\sqrt{(1-p)^2+4px}-(1-p)}{2p}\notag,
\end{align}
where \eqref{eq:ineq3} follows from Lemma \ref{sum-rate}. The
polynomial $f(y)=py^2+(1-p)y-x$ is a convex function of $y$, with the
only positive root at $y = \frac{\sqrt{(1-p)^2+4px}-(1-p)}{2p}$.  So,
if $y < \frac{\sqrt{(1-p)^2+4px}-(1-p)}{2p}$, then $f(y)<0$. Hence,
$\left[(1-p)+p\bar{\lambda}(\epsilon,x)\right]\bar{\lambda}(\epsilon,x)-x<0$
and the density evolution equation converges, as long as
$x\geq\frac{1}{N}$. So, the probability of erasure is upper bounded by
$1/N$.\par
Note that $\int_0^1\rho^{(N)}(x)\,\text{d}x$ is a monotonically
increasing sequence, upper bounded by $1-\frac{p}{2}$. So, in the
limit of infinite blocklengths the design rate is given by 
\begin{equation*}
  R =
  \lim_{N\to \infty}\frac{\int_0^1\lambda(x)\,\text{d}x}{\int_0^1\rho^{(N)}(x)\,\text{d}x}
  = \frac{(1-\epsilon)(1-e^{-\alpha})}{\mu+ (1-\frac{p}{2})}.
\end{equation*}
